\documentclass[11pt,a4paper]{article}

\usepackage{hyperref}

\usepackage[fleqn]{amsmath}
\usepackage{amsthm,amssymb,epsfig,latexsym,euscript}
\usepackage[numbers,square,sort&compress]{natbib}

{\theoremstyle{remark}
}
{\theoremstyle{remark}
}
\makeatletter
\@addtoreset{equation}{section}
\makeatother

\newcommand{\Int}{\int\limits}

\newcommand{\sul}{\sum\limits}
\newcommand{\pl}{\prod\limits}
\newcommand{\dd}{\mathrm{d}}

\newcommand{\bra}[1]{\langle\,#1\,|}
\newcommand{\ket}[1]{|\,#1\,\rangle}
\newcommand{\braket}[2]{\ensuremath{\langle\, #1 \mid  #2\, \rangle }}
\newcommand{\moy}[1]{\langle\,#1\,\rangle}

\newcommand{\Rset }{{\mathbb R }}
\newcommand{\Zset }{{\mathbb Z}}

\newcommand{\mc}[1]{\ensuremath{\mathcal{#1}}}

\def\veps{\varepsilon}
\def\eps{\epsilon}

\def\la{\lambda}

\def\om{\omega}

\begin{document}

\begin{flushright}
LPENSL-TH-06/12
\end{flushright}

\vspace{24pt}

\begin{center}
\begin{LARGE}
{\bf Form factor approach to dynamical correlation functions in critical models}
\end{LARGE}

\vspace{50pt}

\begin{large}
{\bf N.~Kitanine}\footnote[1]{ IMB, UMR 5584 du CNRS, Universit\'e
de Bourgogne, France, Nikolai.Kitanine@u-bourgogne.fr},~~
{\bf K.~K.~Kozlowski}\footnote[2]{ IMB, UMR 5584 du CNRS, Universit\'e
de Bourgogne, France, Karol.Kozlowski@u-bourgogne.fr},~~
{\bf J.~M.~Maillet}\footnote[3]{ Laboratoire de Physique, UMR 5672
du CNRS, ENS Lyon,  France,
 maillet@ens-lyon.fr},\\
{\bf N.~A.~Slavnov}\footnote[4]{ Steklov Mathematical Institute,
Moscow, Russia, nslavnov@mi.ras.ru},~~
{\bf V.~Terras}\footnote[5]{ Laboratoire de Physique, UMR 5672 du
CNRS, ENS Lyon,  France, veronique.terras@ens-lyon.fr}
\end{large}

\vspace{1cm}

\today

\end{center}

\vspace{1cm}

\begin{abstract}
We develop a  form factor  approach to the study of dynamical correlation functions of quantum integrable models in the critical regime. As an example, we consider the quantum  non-linear Schr\"{o}dinger model. We derive long-distance/long-time asymptotic behavior of various two-point functions of this model.  We also compute edge exponents and amplitudes characterizing the power-law behavior of dynamical response functions on the particle/hole excitation thresholds.
These last results confirm predictions based on the non-linear Luttinger liquid method.
Our results rely on a first principles derivation, based on the microscopic analysis of the  model, without invoking, at any stage, some correspondence with a continuous field theory. Furthermore, our approach only makes use of certain general properties of the model, so that it should be applicable, with possibly minor modifications, to a wide class of (not necessarily integrable) gapless one dimensional Hamiltonians.
\end{abstract}

\vspace{1cm}

\section{Introduction}
\label{sec-intro}

Recently \cite{KitKMST11b}, we developed a new method to obtain large distance asymptotic behavior of correlation functions in critical models from their form factor expansion in finite volume. It originates in a careful analysis of the form factor series for large but finite systems that become critical in the infinite size limit. In such a limit, form factors of local operators can be shown to scale to zero as a non-trivial power law of the system size related in particular to their conformal dimension \cite{KitKMST09c,KitKMST11a}. In the asymptotic regime (at large distances) and for large (but finite) system size, the form factor series localizes to its critical part. Namely, it can be shown that only a subset\footnote{This subset contains nevertheless an infinite number of states in the thermodynamic limit.} of the states  contributes to the leading asymptotic behavior of the correlation function. We called these states and their associated  form factors {\em critical}. It turns out that the corresponding series over these critical states and form factors can be summed up thanks to a beautiful (purely combinatorial) multiple sum formula (see \cite{KitKMST11b}). It is worth mentioning that this summation formula has a deep group theoretical origin in the representation theory of the infinite permutation group \cite{KerOV93,Ker2000,BorO2000,BorO2000a,Oko01,BorO2001,Ols2003}. In the context of critical models, it gives the explicit correspondence between the microscopic dynamics and its critical (conformal) limit. Moreover, for integrable models, this result agrees with the direct asymptotic analysis of the correlation function performed from their multiple integral representations obtained in the algebraic Bethe ansatz framework \cite{KitMT99,KitMT00,KitMST02a,KitMST05a,KitMST05c,KitKMST07} and using rather sophisticated Riemann--Hilbert techniques \cite{KitKMST09,KitKMST09b}.

The result of this form factor series analysis is to provide, in a comparatively rather elementary way, the full leading  asymptotic expansion of correlation functions in terms of the distance between local operators. Namely, for each oscillating harmonics quantized in terms of  the Fermi momentum, it leads to the corresponding exact critical exponent together with the associated (non universal) amplitude. In particular, this approach succeeds to confirm, from a computation based on  first principles, the full Luttinger liquid or conformal field theory (CFT) predictions  \cite{LutP75,Hal80,Hal81a,Hal81b,BelPZ84,Car84,Aff85,BloCN86,Car86,AlcBB88,WoyE87} for the asymptotic behavior of correlation function in terms of the distance. It also  provides in addition the corresponding amplitudes written in terms of specific and properly normalized form factors of local operators; this last information, being model dependent, is naturally out of reach of Luttinger liquid theory or CFT. It should be emphasized here that the results obtained in \cite{KitKMST11b} using this method do not rely on any conjectural correspondence of the model with a CFT or Luttinger liquid theory (or even its non-linear version see e.g. \cite{ImaG08,CheP08,ImaG09,ImaG09a,CauGIS10,ImaSG11} and references therein).

The aim of the present article is to apply this new method to the study of dynamical correlation functions and associated structure factors of critical models. As it is well known, the dynamical structure factors (Fourier transforms of the dynamical correlation functions) give the response functions under external perturbations of the system \cite{VanHo54,VanHo54a,Kub57,KubTH85}. As such they can be measured experimentally, see for examples   \cite{VanHo54,VanHo54a,NagTCPS91,ClaSSBDJ02,LakTFN05,LakTNTPRSKB10,KimKROEMUTMSK06,KonKKKZBSBSCPMSKK10,StaCGIGPK99,CleFFFI09,DaoGDSC07}. It is therefore of immediate importance to design efficient methods to determine the values of these functions. Although the exact and  analytic computation of the structure factors in integrable models throughout the full Brillouin zone is for the moment out of reach of our method (see however the method initiated in  \cite{CauM05,CauHM05,PerSCHMWA06,PerSCHMWA07,CauCS07,Cau09} for a combined analytic and controlled numerics that fits extremely well the actual experimental datas), we will provide exact information for these functions in two situations of interest: the large distance and long time asymptotic behavior of correlation functions and the behavior of the corresponding structure factors near the edges of the spectrum.

Let us be slightly more precise about the content of the present paper.
To implement our method we consider here the 1D Bose gas (or quantum non-linear Schr\"odinger model, QNLS) with delta function interaction. The QNLS model describes the evolution of quantum Bose fields $\Psi(x,t)$, $\Psi^\dagger(x,t)$, subject to canonical equal-time commutation relations, with Hamiltonian
\begin{equation}\label{Ham}
  H=\Int_0^L \bigl( \partial_x\Psi^\dagger\, \partial_x\Psi
                                + c\Psi^\dagger\Psi^\dagger\Psi\Psi-h\Psi^\dagger\Psi  \bigr)\, \dd x.
\end{equation}
Here $L$ denotes the size of the model, and we impose periodic boundary conditions.  We restrict our study to the case of the repulsive regime $c>0$ in the presence of a positive chemical potential $h>0$. In such a case, the Bethe ansatz provides the full description of the spectrum in terms of particles and holes excitations \cite{LieL63,Lie63}.
Since the Hamiltonian $H$ commutes with the number of particle operator, it can be diagonalized separately  in every sector with a fixed number $N$ of particles. In each of these sectors, the model is equivalent to the $N$-body gas of bosons with $\delta$-like interactions that was studied  in~\cite{LieL63,Lie63}.

For this model we study various critical properties associated with the simplest zero-temperature time-dependent correlation functions in the thermodynamic limit $L,N\to\infty$ at $N/L$ fixed. The latter correspond to the expectation values, in the ground state $\ket{\psi_g}$, of products of the form
\begin{equation}\label{def-2pt}
  \moy{\mc{O}^\dagger(x,t)\,\mc{O}(x',t')}
   \equiv \frac{\bra{\psi_g}\, \mc{O}^\dagger(x,t)\,\mc{O}(x',t')\,\ket{\psi_g}}{\braket{\psi_g}{\psi_g}},
\end{equation}
where $\mc{O}(x,t)\equiv e^{it H}\mc{O}(x,0)  e^{-itH}$ stands either for one of the Bose fields $\Psi(x,t)$ or $\Psi^\dagger(x,t)$, or for the local density $j(x,t)\equiv\Psi^\dagger(x,t)\Psi(x,t)$. Due to the translation invariance of the model  we set below
$x'=0$ and $t'=0$.

The  large-distance long-time asymptotic  behavior of these two-point functions (i.e. their behavior at $t\to\infty$, $x\to\infty$ with $x/t$ fixed) was  studied recently  in \cite{KozT11,Koz11}  via the
Riemann--Hilbert technique (see also \cite{MccPS83,MulS84,Sla90,ItsIKV91,KojKS97,ItsS99,Sla99,Gan04,GanK01,GanS03,GanS06} for related works).  It was shown that, for certain correlation functions (in particular for the correlation functions of the densities), the leading asymptotic behavior
cannot be inferred using a  CFT-based method \cite{BelPZ84,Aff85,BloCN86,Car84,Car86}.
The reason is that  a correspondence with a CFT can only grasp effects taking their origin in the linear part of the spectrum.
However, in the presence of time and beyond the $x\gg t$ limit, there exists a saddle-point of the oscillating phase (see below)
which is located in the non-linear part of the spectrum.
The form factor approach developed in the present paper allows us to show that this saddle point gives non-trivial contributions to the asymptotic expansion of two-point functions. In particular,  the form factor series localizes and again it can be computed  using the same multiple sum formula as in \cite{KitKMST11b}.

The problem of the characterization of the behavior of dynamical response functions on the particle/hole
excitation thresholds  was recently considered in \cite{ImaG08,CheP08,ImaG09,CauGIS10} within the non-linear Luttinger liquid approach in relation to the Fermi edge singularity effect, see e.g.  \cite{Mah67,NozD69,Mah81,OthT90,Noz97,GogNT98}.
The form factor method used in the present paper allows us to derive the edge exponents and the associated  amplitudes for the 1D Bose gas
directly from the algebraic Bethe ansatz study of the microscopic model, hence from a first principles based derivation. Again, in this case, one can show that, for the leading behavior of the structure factor at threshold, the form factor series localizes and that the corresponding {\em quasi-critical}  sum  (see Section~\ref{sec-ff-exp} and Section~\ref{sec-asympt}) can be computed  thanks in particular to the multiple sum formula used in \cite{KitKMST11b}.
It appears that the correspondence of our results derived directly from the full form factor series with the one obtained through the non-linear Luttinger liquid model with one impurity relies on the rather non-trivial sub-additivity property of the spectrum on the particle and hole threshold lines. Namely, if this sub-additivity property had failed, the form factor series would have localized in a different way and the simple one-impurity picture used in the non-linear Luttinger liquid approach would be invalid.

The article is organized as follows. We give in Section~\ref{sec-ff-exp} a detailed description of the excited states of the model, which allows us to formulate more precisely the main steps of our method of summation over form factors.
We then successively apply this approach in the two cases mentioned above:
to the calculation of the long-time and long-distance asymptotic behavior of two-point functions in Section~\ref{sec-asympt},
and to the computation of the edge exponents of dynamical response functions in Section~\ref{sec-edge}.
Several auxiliary formulae are gathered in two appendices.

\section{Form factor expansion of correlation functions}
\label{sec-ff-exp}

The form factor series representation of the two-point function \eqref{def-2pt} is obtained by inserting,  in between the local operators $\mc{O}^\dagger(x,t)$ and $\mc{O}(0,0)$, a complete set of eigenstates $\ket{\psi'}$ of the Hamiltonian:
\begin{align}
  \moy{\mc{O}^\dagger(x,t)\,\mc{O}(0,0)}
  &=\sum_{\ket{\psi'}}\frac{ \bra{\psi_g}\, \mc{O}^\dagger(x,t)\,\ket{\psi'}\, \bra{\psi'}\,\mc{O}(0,0)\,\ket{\psi_g}}{\braket{\psi_g}{\psi_g}\, \braket{\psi'}{\psi'}}
    \nonumber\\
   &=\sum_{\ket{\psi'}} e^{-it {\mc{E}}_{\mathrm{ex}}+ix  {\mc{P}}_{\mathrm{ex}} }
   \left|\frac{\bra{\psi'}\,\mc{O}(0,0)\,\ket{\psi_g}}{\|\psi_g\|\|\psi'\|}\right|^2.
    \label{ff-series}
\end{align}
In \eqref{ff-series}, ${\mc{E}}_{\mathrm{ex}}$ (resp. ${\mc{P}}_{\mathrm{ex}}$) represents the relative excitation energy (resp. momentum) of the excited state $\ket{\psi'}$ with respect to the
ground state $\ket{\psi_g}$.

There are several advantages of considering a series such as \eqref{ff-series}.
The first one is that the whole time and space dependence is gathered within a single phase factor.
The second one is that there exist explicit determinant representations for the matrix elements
$\bra{\psi'}\,\mc{O}(0,0)\,\ket{\psi_g}$ (see \cite{KojKS97,KozT11}) in finite volume.
The main problem is to sum up the series \eqref{ff-series}.
In this section, we explain our general strategy to perform such a summation in the large size limit using the precise description of excited states given by Bethe ansatz.

\subsection{Description of the spectrum and classes of excited states}

The form factor approach requires a precise description of the spectrum of the Hamiltonian, namely of the ground state and  the excited states over which we sum up in \eqref{ff-series}.
Within the algebraic Bethe ansatz framework, the ground state $\ket{\psi_g}\equiv\ket{\psi(\{ \widehat{\la} \})}$ of \eqref{Ham} is parameterized by a set of
real Bethe parameters $ \widehat{\la}_1,\ldots , \widehat{\la}_N$ solution of the following system of logarithmic Bethe equations:
\begin{equation}\label{Bethe-gr}
  L \, p_0( \widehat{\la}_j)+\sum_{k=1}^N\theta(  \widehat{\la}_j-  \widehat{\la}_k) = 2\pi \Big(j-\frac{N+1}{2}\Big),\qquad j=1,\ldots,N.
\end{equation}
Here the functions $p_0(\la)$ and $\theta(\la)$ correspond respectively to the bare momentum and bare phase of the quasi-particles which, for the QNLS model, are given as
\begin{equation}\label{bare-m-p}
 p_0(\la)=\la,
 \qquad
 \theta(\la)= i\log \Big(\frac{ic+\la}{ic-\la}\Big).
\end{equation}
The number $N$ of Bethe parameters parameterizing the physical ground state of the model is fixed by the chemical potential.
In the thermodynamic limit $L\to\infty$, the ratio $N/L$ has a finite limit $D$, and the ground state Bethe roots $\widehat\la_j$ densely fill a symmetric interval $[-q,q]$ of the real axis (the Fermi zone) with a density function $\rho(\la)$.

Excited states $\ket{\psi(\{  \widehat{\mu}_\ell\})}$ of \eqref{Ham} are parameterized by sets of solutions $  \widehat{\mu}_{\ell_1},\ldots,  \widehat{\mu}_{\ell_{N'}}$ of logarithmic Bethe equations corresponding to different choices of integers $\ell_1<\ell_2<\cdots<\ell_{N'}$:
\begin{equation}\label{eq-Bethe}
 L\, p_0(  \widehat{\mu}_{\ell_j})+\sum_{k=1}^{N'}\theta(  \widehat{\mu}_{\ell_j} -  \widehat{\mu}_{\ell_k})  =2\pi\Big(\ell_j-\frac{N'+1}{2}\Big),
 \qquad j=1,\ldots,N'.
\end{equation}
For the QNLS model, all roots of \eqref{eq-Bethe} are real, and are completely determined by the chosen set of integers $\ell_j$ (so that we have labeled them accordingly).
The state parameterized by the solutions of \eqref{eq-Bethe} with
$\ell_j=j$, $j=1,\dots,N'$, is called the ground state in the $N'$-sector (it differs from the physical ground state of the model if $N'\neq N$).

Hence, the sum over all excited states $\ket{\psi'}$ in \eqref{ff-series} corresponds to a sum over all solutions $\{ \widehat\mu_{\ell_j} \}$ of the logarithmic Bethe equations \eqref{eq-Bethe}, i.e. to a sum over all possible choices of integers $\ell_1<\cdots<\ell_{N'}$ (with $N'=N$ if $\mc{O}(x,t)=j(x,t)$, $N'=N-1$ if $\mc{O}(x,t)=\Psi(x,t)$ and $N'=N+1$ if $\mc{O}(x,t)=\Psi^\dagger(x,t)$).
For a given excited state $\ket{\psi'}\equiv\ket{\psi(\{ \widehat{ \mu}_{\ell_j}\})}$, the relative excitation momentum and energy with respect to the ground state are equal to
\begin{equation}\label{Mom-Ener-finite}
   {\mc{P}}_{\mathrm{ex}}=\sum_{a=1}^{N'} p_0( \widehat{\mu}_{\ell_a})-\sum_{a=1}^N p_0( \widehat{\la}_a),
   \qquad
   {\mc{E}}_{\mathrm{ex}}=\sum_{a=1}^{N'} \veps_0( \widehat{\mu}_{\ell_a})-\sum_{a=1}^N \veps_0( \widehat{\la}_a),
\end{equation}
where $p_0$ is given by \eqref{bare-m-p} and $\veps_0(\mu)=\mu^2-h$.

A convenient way of describing these excited states, in particular in the large size limit, is to use the language of particles and holes:  an excited state with $n$ particles/holes above the ground state in the $N'$-sector corresponds to a solution of \eqref{eq-Bethe} with $\ell_j=j$ except for  $n$ integers $h_1,\dots,h_n\in\{1,\dots,{N'}\}$ for which
$\ell_{h_a}=p_a\in \Zset\setminus\{1,\dots,{N'}\}$.
The
integers $h_a$ correspond to `holes' with respect to the distribution of
integers for the ground state in the $N'$-sector, whereas the
integers $p_a$ correspond to `particles'.
To each choice of such particle/hole integers (or quantum numbers) $p_a$ and $h_a$ one can associate, through the so-called counting function of the corresponding state (see Appendix~\ref{app-therm} for details), sets of particle $\widehat\mu_{p_a}$ and hole $\widehat\mu_{h_a}$ rapidities ($a=1,\ldots ,n$).
In the thermodynamic limit, these particle/hole rapidities tend to some finite values that we denote by $\mu_{p_a}$ and $\mu_{h_a}$.

There are therefore two possible ways of characterizing a given particle/hole excited state: either in terms of the integer quantum numbers $p_a$ and $h_a$ or in terms of the corresponding particle/hole rapidities.
The former will be referred to as a {\em microscopic description} of the excited state, and the latter as a {\em macroscopic description}.
The meaning of this terminology is the following:
for size $L$ large but finite, a finite deviation of the integer $p_a$ (resp. $h_a$), say, $p_a\to p_a+k$ with $|k|\ll N$, leads to a deviation of order $L^{-1}$ for the corresponding rapidity $\widehat{\mu}_{p_a}$ (resp. $\widehat{\mu}_{h_a}$) and deviations of order $L^{-2}$ for all other rapidities.
Therefore, although these two ways of characterizing the particle/hole excited states are equivalent as far as the system size $L$ remains finite, one can find, in the thermodynamic limit, infinitely many microscopic configurations of  the quantum numbers describing an excited state with fixed values $\mu_{p_a}$ and $\mu_{h_a}$ of the macroscopic excitation rapidities.
In particular, the thermodynamic limit of the rapidities of particles (resp. holes) may coincide, while the quantum numbers $p_a$ (resp. $h_a$) are always pair-wise distinct.

We now define classes of excited states that play a role in the process of summation of the series \eqref{ff-series}. We say that two excited states belong to the same class $\mathbf{P}$ if:

\begin{enumerate}
\item[(i)] {\it they have the same excitation momentum ${\mc{P}}_{\mathrm{ex}}$ and energy ${\mc{E}}_{\mathrm{ex}}$ in the thermodynamic limit;}

\item[(ii)] {\it they have the same number of particles and holes separated from the Fermi boundaries\footnote{A particle (resp. a hole) is said to be separated from the Fermi boundaries if the corresponding rapidity $\mu_{p_a}$ (resp. $\mu_{h_a}$) differs from $\pm q$ in the thermodynamic limit. Note that we say nothing here about the number of particles and holes which are {\em not} separated from the Fermi boundaries (except that point (i) should be satisfied).
In particular, the total number of excitations is not fixed.}, with the same rapidities $\mu_{p_a}$ and $\mu_{h_a}$ in the thermodynamic limit.}
\end{enumerate}

Since there is a one to one correspondence between excited states and form factors, the latter can also be gathered into
classes in the sense of the above definition.

It follows from this definition that, within a given class, the excited states (for $N,L$ large) differ from each others not only by $1/L$ excitations around the (macroscopic) particle $\mu_{p_a}$
or hole $\mu_{h_a}$ rapidities away from the Fermi boundaries,
but also by the possible creation of different numbers of particles and holes on the right
or left Fermi boundaries $\pm q$.
Being on the macroscopic level, one can in fact distinguish states of the same class only on the basis of their total number of particle-hole excitations on each of the Fermi
boundaries. Two states of a given class with the same number of such particle-hole excitations can nevertheless be distinguished at the microscopic level by their quantum numbers $p_a$ and $h_a$.

In \cite{KitKMST11b}, we dealt with particular classes of excited states (resp. form factors) which were called critical states (resp. critical form factors).
In such classes, all particles and holes are accumulated on the Fermi boundaries.
In the present paper we also consider classes of excited states with several particles (holes) separated from the Fermi boundaries.
We call such excited states (resp. form factors) {\it quasi-critical states} (resp. {\it quasi-critical form factors}).

\subsection{Summation process}

The main idea of our method is to use our partitioning of excited states into classes so as to split the summation of the form factor series \eqref{ff-series} into two steps:
 \begin{equation}\label{sum-class}
 \sum_{\ket{\psi'}} e^{-it {\mc{E}}_{\mathrm{ex}}+ix  {\mc{P}}_{\mathrm{ex}} }
   \left|\frac{\bra{\psi'}\,\mc{O}(0,0)\,\ket{\psi_g}}{\|\psi_g\|\|\psi'\|}\right|^2=
   \sum_{\mathbf{P}} \sum_{\scriptscriptstyle\ket{\psi'}\in\mathbf{P}} e^{-it {\mc{E}}_{\mathrm{ex}}+ix  {\mc{P}}_{\mathrm{ex}} }
   \left|\frac{\bra{\psi'}\,\mc{O}(0,0)\,\ket{\psi_g}}{\|\psi_g\|\|\psi'\|}\right|^2.
 \end{equation}
\begin{enumerate}
\item In \eqref{sum-class} we first compute the sum of form factors of a given class  $\mathbf{P}$, using the microscopic description of the excited states.
We should in particular take into account, in the large-size representation of the form factors (see \cite{KitKMST11a,KozT11,Koz11}), the part that non-trivially depends on the quantum numbers.
We have argued in \cite{KitKMST11b} that this part is a quite universal object that, to a large extend,  is model-independent. Due to its specific form (see Section~\ref{sec-asympt} for details), the sum of quasi-critical form factors of a given class can be calculated in the large-size limit using the multiple sum formula of  \cite{KitKMST11b}.
\item The second step consists in summing up the contributions from different classes. According to the problem we consider (see Sections~\ref{sec-asympt} and \ref{sec-edge}), it is possible to argue that only certain very specific classes $\mathbf{P}$ will contribute, at leading order, to the critical behavior we are interested in.
At this stage of the calculation, it becomes enough, for the cases we are dealing with, to use the macroscopic description of the excited states, and to replace the sum over classes by integrals over the macroscopic particle/hole rapidities $\mu_{p,h}$. Moreover, the obtained integrals are localized around very specific particle/hole rapidities $\mu_{p,h}$.
\end{enumerate}

Let us be slightly more specific.
It was shown in \cite{Sla90,KitKMST09c,KitKMST11a} that  particle-hole form factors scale to zero in the thermodynamic limit.
More precisely, if the system size $L$ goes to infinity, they decrease with some negative power of $L$, which depends on the corresponding particle/hole rapidities:
 \begin{equation}\label{Gen-ff}
 \left| \frac{\bra{\psi'}\,\mc{O}(0,0)\,\ket{\psi_g}}
           { || \psi_g || \; ||\psi'|| } \right|^2=L^{-\theta( \{ \mu_{p,h}\})}
           A ( \{ \mu_{p,h}\}| \{p,h\}  ).
 \end{equation}
Here we have pointed out  that the finite amplitude $A$ depends both on the quantum numbers $\{p,h\}$ and on the particle/hole rapidities $\{ \mu_{p,h}\}$. This notation stresses some important properties of the amplitude.
Namely, if a particle or a hole is separated from the Fermi boundaries by a finite distance, then a finite change of the corresponding quantum number results into a variation of order $L^{-1}$ of the amplitude $A$; in this case, one can say that $A$ depends on the macroscopic rapidities $\mu_{p,h}$ rather than on the quantum numbers or that, in other words, it is enough to use a macroscopic description.
On the contrary, for a particle or a hole on the Fermi boundary, (i.e. such that $\widehat{\mu}_{p_a}$ or $\widehat{\mu}_{h_a}$ tends to $\pm q$ in the thermodynamic limit), then a finite deviation $p_a\to p_a+k$ (resp. $h_a\to h_a+k$) of the corresponding microscopic quantum number leads to a finite variation of the amplitude; in this case the microscopic description should be used.
This property of the amplitude becomes extremely important when we sum up form factors of a given class because this amounts to summing up excitations at the Fermi boundaries.

Calculating the sum of quasi-critical form factors we should take into account $L^{-1}$-cor\-rec\-ti\-ons to the excitation momentum and energy \cite{KitKMST11b}. For $L$ large but finite the latter can be presented as
 \begin{equation}\label{Mom-Ener0}
   {\mc{P}}_{\mathrm{ex}}={\mc{P}}^{(0)}_{\mathrm{ex}}+{\mc{P}}^{(1)}_{\mathrm{ex}},\qquad
   {\mc{E}}_{\mathrm{ex}}={\mc{E}}^{(0)}_{\mathrm{ex}}+{\mc{E}}^{(1)}_{\mathrm{ex}},
\end{equation}
where the thermodynamic limits ${\mc{P}}^{(0)}_{\mathrm{ex}}\equiv {\mc{P}}^{(0)}_{\mathrm{ex}}(\{\mu_{p,h}\})$ and ${\mc{E}}^{(0)}_{\mathrm{ex}}\equiv {\mc{E}}^{(0)}_{\mathrm{ex}}(\{\mu_{p,h}\})$,
\begin{equation}\label{Mom-Ener}
   {\mc{P}}^{(0)}_{\mathrm{ex}}=\sum_{a=1}^{n}\bigl[ p(\mu_{p_a})- p(\mu_{h_a})\bigr],
   \qquad
   {\mc{E}}^{(0)}_{\mathrm{ex}}=\sum_{a=1}^{n}\bigl[ \varepsilon(\mu_{p_a})- \varepsilon(\mu_{h_a})\bigr],
\end{equation}
are given in terms of the dressed momentum \eqref{dmom} and the dressed energy \eqref{denergy} of the particles and the holes at the thermodynamic limit,
whereas the corrections ${\mc{P}}^{(1)}_{\mathrm{ex}}\equiv {\mc{P}}^{(1)}_{\mathrm{ex}} ( \{ \mu_{p,h}\}| \{p,h\}  )$ and ${\mc{E}}^{(1)}_{\mathrm{ex}}\equiv {\mc{E}}^{(1)}_{\mathrm{ex}} ( \{ \mu_{p,h}\}| \{p,h\}  )$ are of order $L^{-1}$ (see \eqref{dec-Pex}, \eqref{dec-Eex} for their explicit representations).
Note that  ${\mc{P}}^{(0)}_{\mathrm{ex}}$ and ${\mc{E}}^{(0)}_{\mathrm{ex}}$ can be described on a macroscopic level and that they take the same value for all quasi-critical states in a given class.
On the contrary, ${\mc{P}}^{(1)}_{\mathrm{ex}}$ and ${\mc{E}}^{(1)}_{\mathrm{ex}}$ explicitly depend on quantum numbers, and therefore on the specific representative of the class that we consider.

In principle, so as to be consistent with the above degree of approximation, one should also take care about the $L^{-1}$-corrections to the large-size representation \eqref{Gen-ff}.
However, in the problems we deal with (long-time long-distance asymptotic behavior and calculation of the edge exponents), such corrections only produce subleading contributions to the results, and we can therefore neglect them.
In particular, one can consider the power $L^{-\theta( \{ \mu_{p,h}\})}$ as a common prefactor $L^{-\theta_{\mathbf{P}} }$ for a given class $\mathbf{P}$ of quasi-critical form factors.

Therefore, restricting the form factor series \eqref{sum-class} to the sum over quasi-critical states belonging to a given class $\mathbf{P}$, we obtain, at our order of approximation in $L^{-1}$,
 \begin{multline}\label{sum-class1}
  \sum_{\scriptscriptstyle\ket{\psi'}\in\mathbf{P}} e^{-it {\mc{E}}_{\mathrm{ex}}+ix  {\mc{P}}_{\mathrm{ex}} }
   \left|\frac{\bra{\psi'}\,\mc{O}(0,0)\,\ket{\psi_g}}{\|\psi_g\|\|\psi'\|}\right|^2\\
   =   L^{-\theta_{\mathbf{P}} }
        e^{-it {\mc{E}}_{\mathrm{ex}}^{(0)}+ix  {\mc{P}}_{\mathrm{ex}}^{(0)} }
   \sum_{ \{p\},\{h\} } e^{ {ix}{\cal P}_{\mathrm{ex}}^{(1)}(\{p,h\})-{it}{\cal E}_{\mathrm{ex}}^{(1)}(\{p,h\})}
                                  A ( \{ \mu_{p,h}\}| \{p,h\}  ).
 \end{multline}
Here the values of the rapidities $\mu_{p,h}$ in the arguments of the amplitude $A$ are fixed.

In \cite{KitKMST11b}, we dealt with the sum \eqref{sum-class1} in the case ${\mc{E}}_{\mathrm{ex}}^{(0)}=0$ and $t=0$.
We have in particular shown there that, in this case, the computation of the sum eventually produces a factor $L^{\theta_{\mathbf{P}} }$ precisely compensating the prefactor $L^{-\theta_{\mathbf{P}} }$,
thus ensuring that the final result has a finite thermodynamic limit.
It means that the sum over excitations on the Fermi surface provides an effective dressing of the original form factors.

In our present case, with more general classes such that ${\mc{E}}_{\mathrm{ex}}^{(0)}\ne 0$, this dressing compensates the vanishing pre-factor $L^{-\theta_{\mathbf{P}} }$ only partly.
However, the remaining negative power of $L$ disappears when we sum up the contributions of different classes: it can actually be absorbed into the integration measure when one replaces the summation over classes in \eqref{sum-class} by an integration over the particle/hole rapidities separated from the Fermi boundaries.

To achieve this second step, and to evaluate the two-point correlation functions, it finally remains to compute these integrals over $\{\mu_{p,h}\}$.
Whereas their explicit analytic evaluation is hardly possible in the general case (this is due to the very complicated dependence of the amplitude $A$ on the rapidities $\{\mu_{p,h}\}$,  see e.g. \cite{KozT11}), the situation can be simplified if, for some reason, these integrals are localized around one point (or several points): the amplitude can then be replaced by a constant, namely by its value at the point of localization.
In the examples discussed below, we are precisely dealing with such a case.

\section{Long-time and large-distance asymptotic behavior of the two-point functions}
\label{sec-asympt}

In this section we study the long-time and large-distance  ($x\rightarrow\infty$, $t\rightarrow \infty$,  with $x/t$ being kept constant) asymptotic behavior of the time-dependent two-point correlation functions, using the form factor summation process sketched in the previous section.

\subsection{General scheme}
\label{sec-general}

Following the same method as for the time-independent case \cite{KitKMST11b}, we first need to identify, within the series \eqref{ff-series}, the relevant form factors contributing to the leading asymptotic
of a given oscillating harmonics.

In the thermodynamic limit, for states with $n$ particles and $n$ holes, the oscillating phase can be presented in the form:
\begin{equation}
   -it{\mc{E}}_{\mathrm{ex}}^{(0)}+ix{\mc{P}}_{\mathrm{ex}}^{(0)}
   =-it\sum_{a=1}^{n}\bigl[u(\mu_{p_a})-u(\mu_{h_a})\bigr],
\label{phase-u}
\end{equation}
where
\begin{equation}\label{def-u}
  u(\la)= \varepsilon(\la)-\frac xt p(\la).
\end{equation}
In the limit $x\rightarrow\infty$, $t\rightarrow \infty$ with $x/t=\mathrm{const}$, the phase factor becomes
rapidly oscillating.
Similarly to what happens for oscillating integrals, we therefore expect that the main contributions to the sum come either from the boundaries of the summation interval (the Fermi boundaries $\pm q$) or from the phase saddle-point. In the equal-time case studied in \cite{KitKMST11b}, there was no phase saddle-point, and  the leading asymptotic contribution was solely issued from states characterized by particle and hole excitations close to the Fermi boundaries.
However, in the present dynamical case, the saddle-point contributions should also be taken into account.

The existence and the uniqueness of the saddle-point of the phase $u(\lambda)$ \eqref{def-u} is closely related to the properties of the  sound velocity  $v(\lambda)$ in the Bose gas:
\begin{equation}\label{v-sound}
v(\lambda)=\frac{\partial\varepsilon}{\partial p}=\frac{\varepsilon'(\la)}{p'(\la)}.
\end{equation}
It is believed that this function is strictly monotonic: $v'(\lambda)>0$.
This property can be proved if the Hamiltonian parameters $c$ and $h$ are such that
$c/q(c,h)\gtrapprox 0.6$.
Otherwise, to the best of our knowledge, the corresponding proof is missing up to now (see e.g. \cite{CorDZ09}). The numerical analysis of equations \eqref{dmom} and
\eqref{denergy} does nevertheless confirm that $v'(\lambda)>0$ for arbitrary values of $c$ and $h$,
and we therefore assume this property.
It also follows from equations \eqref{dmom} and  \eqref{denergy} that $v(\lambda)\to\pm\infty$
as $\lambda\to\pm\infty$. This ensures that the phase $u(\lambda)$ has a unique saddle point $\lambda_0$ determined by the equation $v(\lambda_0)=x/t$.
Moreover, the monotonicity of $v(\lambda)$ associated
to the fact that $p'(\lambda)>0$ (\textit{cf} \eqref{dmom} ) also ensures that  $u''(\lambda_0)>0$.

Introducing the Fermi velocity $v_{{}_F}=v(q)=-v(-q)$, we can define two asymptotic regimes\footnote{
The analysis of the asymptotic behavior for $x/t=\pm v_{{}_F}$ goes way beyond the scope of this paper.}:
\begin{itemize}
\item the space-like regime ($|x/t |>v_{{}_F}$) which corresponds to a saddle-point  outside of the Fermi zone $|\la_0|>q$;
\item the time-like regime  ($|x/t| <v_{{}_F}$)  which corresponds to a saddle-point  inside of the Fermi zone $|\la_0|<q$.
\end{itemize}

We now describe the excited states (and the corresponding form factors) giving rise to the leading
asymptotic terms of the correlation functions.
In the space-like (resp. time-like) regime, the main contribution obviously comes from form factors associated to states for which the particles (resp. holes) which are separated form the Fermi boundaries (if any) are close to the saddle-point $\lambda_0$.
In other words, a contributing state with $n$ particle-hole excitations in the space-like (resp. time-like) regime is such that there are $n_0$ particles (resp. holes) with rapidities $\{\mu_{s_a}\}$ lying in a small vicinity of the saddle point $\lambda_0$, the other $n-n_0$ particles (resp. holes), as well as the total number $n$ of holes (resp. particles) being on the Fermi boundaries $\pm q$, with $n_0 \ge 0$.
To describe such states, it is convenient to introduce a parameter $\tau$ such that  $\tau=1$ in the space-like regime  and  $\tau=-1$ in the time-like regime.
Let also $n_p^\pm$ be the number of particles with rapidity $\pm q$, and  $n_h^\pm$ be the number of holes with rapidities $\pm q$.
Then, in both regimes, the above-introduced numbers counting the amount of particles (resp. holes)
all sum up to the total number of excitations:
\begin{equation}
     n_p^+ + n_p^- +  \frac{(1+\tau)}{ 2 }  n_0 =n_h^+ + n_h^-  +  \frac{(1-\tau)}{ 2 }  n_0=n .
\label{TotNP}
\end{equation}
Let $\ell$ refer to the right particle/hole discrepancy  number $\ell=n_p^{+} - n_h^{+}$.
Due to the constraints \eqref{TotNP}, this means that the left particle/hole discrepancy  number can be written as  $n_p^{-} - n_h^{-}=-\tau n_0-\ell$.

The states described above form a class of what we call {\em quasi-critical} states. This class is characterized by the integer numbers $\ell$, $\tau$ and $n_0$, and by the rapidities $\mu_{s_1},\dots, \mu_{s_{n_0}}$ localized in a small vicinity of $\lambda_0$ with $n_0 \ge 0$. If these parameters are fixed, then, in the thermodynamic limit, all states within this class have the same energy and momentum, namely
 \begin{equation}\label{En-Mom1}
   {\mc{E}}_{\mathrm{ex}}^{(0)}=\tau\sum_{a=1}^{n_0}\varepsilon(\mu_{s_a}),\qquad
   {\mc{P}}_{\mathrm{ex}}^{(0)}=\tau\sum_{a=1}^{n_0}p(\mu_{s_a})+(2\ell+\tau n_0)k_{{}_F},
   \end{equation}
where $k_{{}_F}=p(q)$ is the Fermi momentum. We refer to this class of quasi-critical states (and form factors) as the $\mathbf{P}_{\tau n_0,\ell}$ class.

As mentioned in Section~\ref{sec-ff-exp}, all the form factors of the $\mathbf{P}_{\tau n_0,\ell}$ class scale to zero with the same exponent that we denote by $\theta_{\tau n_0,\ell}$.
The latter  was computed for generic particle-hole form factors in \cite{KozT11,KitKMST11a}.
It was shown there that it can be expressed in terms of the values of the shift function (see \eqref{shift-ph}) on the boundaries of the Fermi zone:
\begin{equation}\label{crit-theta}
  \theta_{\tau n_0,\ell}=(F_{\tau n_0,\ell}^+ +\ell)^2+(F_{\tau n_0,\ell}^- +\ell+\tau n_0)^2+n_0,
\end{equation}
with
\begin{equation}\label{shift1}
  F_{\tau n_0,\ell}^-= F_{\tau n_0,\ell}(-q), \quad F_{\tau n_0,\ell}^+\equiv F_{\tau n_0,\ell}(q)+\Delta N.
\end{equation}
The integer $\Delta N=N'-N$ is $0$ for the correlation function of densities and $\pm1$ for the correlation functions of fields,
and the shift function $F_{\tau n_0,\ell}$ associated with the $\mathbf{P}_{\tau n_0,\ell}$ class takes the form
\begin{equation}\label{shift2}
  F_{\tau n_0,\ell}(\lambda)
   =  - \Delta N \Big[   \frac{Z(\la)}{2} + \phi(\lambda,q)  \Big]
       -  \ell \phi(\lambda,q)   +   (\ell + \tau n_0)  \phi(\lambda,-q)
       -  \tau \sum_{a=1}^{n_0} \phi(\lambda,\mu_{s_a})
\end{equation}
in terms of the dressed charge $Z(\la)$ and dressed phase $\phi(\la,\nu)$ defined respectively in \eqref{dcharge} and \eqref{dphase}.

In order to fix a quasi-critical state inside the $\mathbf{P}_{\tau n_0,\ell}$ class, we need to introduce the quantum numbers labeling the particles and holes with rapidities equal to $\pm q$.
We use the standard (see \cite{KitKMST11b}) re-parametrization of the original quantum numbers $\{p\}$ and $\{h\}$:
 \begin{equation}\label{spec-p0}
 \begin{array}{ll}
 p_j=p_j^++N',&\mbox{if}\quad \mu_{p_j}=q \, ,\\
 p_j=1-p_j^-,&\mbox{if}\quad \mu_{p_j}=-q \, ,\\
 h_j=N'+1-h_j^+,&\mbox{if}\quad \mu_{h_j}=q \, ,\\
 h_j=h_j^-,&\mbox{if}\quad \mu_{h_j}=-q \, .
 \end{array}
 \end{equation}
These integers allow us to express the amplitude $A$ and the finite-size corrections
to the oscillating phase.

A careful analysis of the finite-size corrections in \eqref{Mom-Ener-finite} enables us to show that the corrections to the momentum and energy in \eqref{Mom-Ener0}
have the form
\begin{align}
 &{\mc{P}}_{\mathrm{ex}}^{(1)}
   =\frac1L\mathcal{P}_{\tau n_0,\ell}
     +\frac{2\pi}L  \Bigg[  \sul_{a=1}^{n_p^+}( p_a^+ -1)+\sul_{a=1}^{n_h^+} h_a^+
     -\sul_{a=1}^{n_p^-} (p_a^- -1) -\sul_{a=1}^{n_h^-} h_a^- \Bigg],
      \label{dec-Pex}\\
 &{\mc{E}}_{\mathrm{ex}}^{(1)}
   =\frac1L\mathcal{E}_{\tau n_0,\ell}
     +\frac{2\pi v_{{}_F}}L \Bigg[  \sul_{a=1}^{n_p^+} (p_a^+ -1)
     +\sul_{a=1}^{n_h^+} h_a^++\sul_{a=1}^{n_p^-} (p_a^- -1) +\sul_{a=1}^{n_h^-} h_a^- \Bigg],
      \label{dec-Eex}
\end{align}
where $\mathcal{P}_{\tau n_0,\ell}$ and $\mathcal{E}_{\tau n_0,\ell}$ are constants (at least at the leading order in $L$) for all states belonging to a given $\mathbf{P}_{\tau n_0,\ell}$ class.

The explicit form of the amplitude $A$ for the quasi-critical form factors  can be obtained following the same lines as for the critical form factors. There is only one subtlety due to the localization of the rapidities $\mu_{s_a}$ in the vicinity of $\lambda_0$. The matter is that, due to the fermionic structure of the excitations, the form factors vanish as soon as two or more rapidities of particles coincide.
Actually, the amplitude $A$ has the form $A=\Delta^2(\{\mu_{s_a}\}) \tilde A$, where
$\Delta(\{\mu_{s_a}\})$ is the Vandermonde determinant of the rapidities $\{\mu_{s_a}\}$. The effective amplitude $\tilde A$ has a non-vanishing
limit as $\mu_{s_a}\to \lambda_0$. Thus, as a first approximation,  we can set all $\{\mu_{s_a}\}$
equal to $\lambda_0$ in the arguments of $\tilde A$, but we should keep the prefactor $\Delta^2(\{\mu_{s_a}\})$ as it is.

So as to express this amplitude, it is convenient, as in the time-independent case, to introduce the simplest form factor of the $\mathbf{P}_{\tau n_0,\ell}$ class.
The corresponding quasi-critical state $\ket{\psi'_{{\tau n_0,\ell}}}$ is defined as follows:
 \begin{itemize}\label{simplest-state}
 \item the $n_0$ rapidities $\mu_{s_a}$ (which are the same for all representatives of the class) are in a small vicinity  $J_{\la_0}$ of the saddle point $\lambda_0$;
 \item the distribution of particles and holes on the right Fermi boundary is as follows:
 if $\ell>0$, there is no hole ($n_h^+=0$) and there are $\ell$ particles ($n_p^+=\ell$) which are characterized by the integers $p_a^+=a$, $a=1,\dots,\ell$;
 if $\ell<0$, there is no particle ($n_p^+=0$) and there are $|\ell |$ holes  ($n_h^+=-\ell$) which are characterized by the integers $h_a^+=a$, $a=1,\dots,-\ell$;
\item the distribution of particles and holes on the left Fermi boundary is as follows:
 if $\ell+\tau n_0<0$, there is no hole ($n_h^-=0$) and there are $-\ell - \tau n_0$ particles ($n_p^-=-\ell - \tau n_0$) with $p_a^-=a$, $a=1,\dots,-\ell-\tau n_0$;
 if $\ell+\tau n_0>0$, there is no particle ($n_p^-=0$)  and there are $ \ell + \tau n_0$ holes ($n_h^-=\ell +\tau n_0$)  labelled by the integers $h_a^-=a$, $a=1,\dots,\ell+\tau n_0$.
 \end{itemize}
Using this representative of the $\mathbf{P}_{\tau n_0,\ell}$ class, we introduce the renormalized form factor
\begin{equation}
\label{basic_ff}
\big|\mathcal{F}^{\mathcal{O}}_{{}_{\tau n_0,\ell}}\big|^2
   =\lim_{L\to\infty}
     \lim_{\mu_{s_a}\to\lambda_0}
     L^{\theta_{\tau n_0,\ell}}\,
     \Delta^{-2}(\{\mu_{s_a}\})\,
     \frac{|\bra{\psi'_{{\tau n_0,\ell}}}\, \mathcal{O}(0,0)\, \ket{\psi_{g}}|^2}
             {\moy{\psi_g \, |\, \psi_g}\,\moy{\psi'_{{\tau n_0,\ell}} \, |\, \psi'_{{\tau n_0,\ell}} } }.
%
\end{equation}
Then, for large $L$ and after setting $\mu_{s_a}=\lambda_0$, the leading order of $\tilde A$ can be written in the following form:
 \begin{multline} \label{ampl}
   \tilde A^{(\tau n_0,\ell)}(\{\lambda_0\}|\{p^\pm,h^\pm\})
   =
   \big|\mathcal{F}^{\mathcal{O}}_{{}_{\tau n_0,\ell}}\big|^2 \,
   \frac{G^2(1+F_{\tau n_0,\ell}^+)\, G^2(1-F_{\tau n_0,\ell}^-)}
          {G^2(1+\ell+F_{\tau n_0,\ell}^+)\, G^2(1-\ell-\tau n_0-F_{\tau n_0,\ell}^-)}
          \\
   \times
  R_{n_p^+,n_h^+}(\{p^+\},\{h^+\} |  F_{\tau n_0,\ell}^+)\,
  R_{n_p^-,n_h^-}(\{p^-\},\{h^-\} |  -F_{\tau n_0,\ell}^-) \, ,
\end{multline}
where $G(z)$ is the Barnes function satisfying  $G(z+1)=\Gamma(z) G(z)$,
and the function $R_{n,n'}$ is
\begin{equation}
\label{def-R}
 R_{n,n'}(\{p\},\{h\}|F)
   = \!\left(\frac{\sin\pi F}\pi\right)^{\!\! 2n'}
      \frac{\prod\limits_{j>k}^n(p_j-p_k)^2\prod\limits_{j>k}^{n'}(h_j-h_k)^2}
             {\prod\limits_{j=1}^n\prod\limits_{k=1}^{n'}(p_j+h_k-1)^2}
     \pl_{k=1}^n \! \frac{\Gamma^2(p_k+F)} {\Gamma^2(p_k)}
     \pl_{k=1}^{n'} \! \frac{\Gamma^2(h_k-F)} {\Gamma^2(h_k)}.
\nonumber
\end{equation}

Thus, the sum of the form factors of the $\mathbf{P}_{\tau n_0,\ell}$ class reads
\begin{multline}
 \sum_{\scriptscriptstyle\ket{\psi'}\in\mathbf{P}_{\tau n_0,\ell}}
  e^{-it {\mc{E}}_{\mathrm{ex}}+ix  {\mc{P}}_{\mathrm{ex}} }
   \left|\frac{\bra{\psi'}\,\mc{O}(0,0)\,\ket{\psi_g}}{\|\psi_g\|\, \|\psi'\|}\right|^2
 =\Delta^2(\{\mu_{s_a} \}) \, L^{-\theta_{\tau n_0,\ell}}\,
   e^{ ix{\cal P}_{\mathrm{ex}}^{(0)}-it{\cal E}_{\mathrm{ex}}^{(0)}  }
   \\
  \times
    \sum_{\{p^\pm\},\{h^\pm\}}\,
   e^{ ix{\cal P}_{\mathrm{ex}}^{(1)}-it{\cal E}_{\mathrm{ex}}^{(1)}  }
   \tilde A^{(\tau n_0,\ell)}(\{\lambda_0\}|\{p^\pm,h^\pm\}).
\label{Cor-fun-ell-pm}
 \end{multline}
Using the explicit representations  \eqref{dec-Pex}, \eqref{dec-Eex} for ${\cal P}_{\mathrm{ex}}^{(1)}$ and ${\cal E}_{\mathrm{ex}}^{(1)}$  and \eqref{ampl} for $\tilde A^{(\tau n_0,\ell)}$
we see that the sum over the quantum numbers $\{p^\pm\},\{h^\pm\}$ can be factorized into a sum over the integers $\{p^+\}, \{h^+\}$ on the one hand, and a sum over the integers $\{p^-\}, \{h^-\}$ on the other hand.
Both of these sums can be computed explicitly via the  identity used in
\cite{KitKMST11b} (see also \cite{KerOV93,Ker2000,BorO2000,BorO2000a,Oko01,BorO2001,Ols2003}):
\begin{multline}
 \sum_{\substack{n,n'\ge 0\\ n-n' = r}}\,
 \sum_{1\le p_1<\cdots<p_{n}<\infty}\,\sum_{1\le h_1<\cdots<h_{n'}<\infty}
 w^{\, \sul_{j=1}^{n}(p_j-1)+\sul_{k=1}^{n'} h_k }    R_{n,n'}(\{p\},\{h\}|F)
     \\
  =\frac{G^2(1+ r +F)}{G^2(1+F)}\,\frac{w^{ r ( r -1)/2}\,}{(1-w)^{(F+ r )^2}}.
  \label{magic-formula}
\end{multline}
In our case one should set
\begin{equation}\label{om-++}
w=e^{\frac{2\pi i }{L}(x-v_{{}_F}t)}, \qquad\ F=F_{\tau n_0,\ell}^+, \qquad r=\ell,
\end{equation}
for the sum over $\{ p^+\}, \{h^+\}$, and
\begin{equation}\label{om---}
  w=e^{- \frac{2\pi i }{L}(x+v_{{}_F}t)}, \qquad F=-F_{\tau n_0,\ell}^-, \quad r=-\ell-\tau n_0,
\end{equation}
for the sum over $\{p^-\}, \{h^-\}$.
Moreover, since the identity \eqref{magic-formula} is valid for $|w|<1$, we should regularize the form factor series by adding a negative imaginary part to the variable $t$, i.e. by replacing $t$ by $t-i0$.
Note that such a regularization is very natural since it increases the convergence of the original form factor series \eqref{ff-series}.
Then, the sum over form factors of the $\mathbf{P}_{\tau n_0,\ell}$ class results into
\begin{align}
   & \sum_{\scriptscriptstyle\ket{\psi'}\in\mathbf{P}_{\tau n_0,\ell}}
    e^{-it {\mc{E}}_{\mathrm{ex}}+ix  {\mc{P}}_{\mathrm{ex}} }
    \left|\frac{\bra{\psi'}\,\mc{O}(0,0)\,\ket{\psi_g}}{\|\psi_g\| \, \|\psi'\|}\right|^2
  =  \big|\mathcal{F}^{\mathcal{O}}_{{}_{\tau n_0,\ell}}\big|^2 \,
      \Delta^2(\{\mu_{s_a} \}) \,  L^{- n_0}\,
      e^{ ix{\cal P}_{\mathrm{ex}}^{(0)}-it{\cal E}_{\mathrm{ex}}^{(0)}  }
           \nonumber \\
  &\quad\times e^{ \frac{\pi i }{L}(x-v_{{}_F}t)\ell(\ell-1)
              -\frac{\pi i }{L}(x+v_{{}_F}t)(\ell+\tau n_0)(\ell+\tau n_0+1)
              +\frac iL(x\mathcal{P}_{\tau n_0,\ell}-t\mathcal{E}_{\tau n_0,\ell})}
         \nonumber  \\
 &\quad\times  L^{-\theta_{\tau n_0,\ell}+ n_0}
     \Big(1-e^{ \frac{2\pi i }{L}(x-v_{{}_F}t)} \Big)^{-(F_{\tau n_0,\ell}^++\ell)^2}
\Big(1-e^{ - \frac{2\pi i }{L}(x+v_{{}_F}t)}\Big)^{-(F_{\tau n_0,\ell}^-+\ell+\tau n_0)^2} .
     \label{Cor-fun-ell-pm-mf}
 \end{align}

It is quite remarkable that now the thermodynamic limit can be taken in the last two lines of this equation. Obviously  the  limit of the second line  is $1$.  Using the explicit form of the exponent $\theta_{\tau n_0,\ell}$ (\ref{crit-theta}) one can easily see that the last line has a finite thermodynamic limit:
\begin{multline}\label{2-line}
 \lim_{L\to\infty}L^{-\theta_{\tau n_0,\ell}+n_0}
    \Big(1-e^{ \frac{2\pi i }{L}(x-v_{{}_F}t)} \Big)^{-(F_{\tau n_0,\ell}^++\ell)^2}
    \Big(1-e^{- \frac{2\pi i }{L}(x+v_{{}_F}t)}\Big)^{-(F_{\tau n_0,\ell}^-+\ell+\tau n_0)^2}
      \\
=\frac{e^{i\frac\pi2[\mathrm{sgn}(x-v_{{}_F}t)(F_{\tau n_0,\ell}^++\ell)^2 -  \mathrm{sgn}(x+v_{{}_F}t)(F_{\tau n_0,\ell}^-+\ell+\tau n_0)^2]}}
{|2\pi(x-v_{{}_F}t) |^{(F_{\tau n_0,\ell}^++\ell)^2}|2\pi(x+v_{{}_F}t)|^{(F_{\tau n_0,\ell}^-+\ell+\tau n_0)^2}}.
\end{multline}
The first line in \eqref{Cor-fun-ell-pm-mf} is proportional to $L^{-n_0}$, but this factor will be absorbed into the integration measure in the second step of the calculation, when we sum up the contributions from different $\mathbf{P}_{\tau n_0,\ell}$ classes.
Therefore in the large $L$ regime,  we can safely take the limit $L\to\infty$ in the last two lines of \eqref{Cor-fun-ell-pm-mf}, which gives us
\begin{multline}
 \sum_{\scriptscriptstyle\ket{\psi'}\in\mathbf{P}_{\tau n_0,\ell}}
    e^{-it {\mc{E}}_{\mathrm{ex}}+ix  {\mc{P}}_{\mathrm{ex}} }
    \left|\frac{\bra{\psi'}\,\mc{O}(0,0)\,\ket{\psi_g}}{\|\psi_g\| \, \|\psi'\|}\right|^2
  =  \big|\mathcal{F}^{\mathcal{O}}_{{}_{\tau n_0,\ell}}\big|^2 \,
      \Delta^2(\{\mu_{s_a} \}) \,  L^{- n_0}\,
      e^{ ix{\cal P}_{\mathrm{ex}}^{(0)}-it{\cal E}_{\mathrm{ex}}^{(0)}  }
   \\
   \times
   \frac{e^{i\frac\pi2[\mathrm{sgn}(x-v_{{}_F}t)(F_{\tau n_0,\ell}^++\ell)^2 -  \mathrm{sgn}(x+v_{{}_F}t)(F_{\tau n_0,\ell}^-+\ell+\tau n_0)^2]}}
{|2\pi(x-v_{{}_F}t) |^{(F_{\tau n_0,\ell}^++\ell)^2}|2\pi(x+v_{{}_F}t)|^{(F_{\tau n_0,\ell}^-+\ell+\tau n_0)^2}}.
     \label{Contr-class}
\end{multline}
Thus, we have calculated the sum over quasi-critical form factors of a single $\mathbf{P}_{\tau n_0,\ell}$ class.
As we have seen, most of the singularity (apart from the $L^{-n_0}$ factor) of the form factors has been absorbed by this process.
In fact, such a summation can be interpreted as a dressing of the original bare simplest form factor of the class by a cloud of  particle-hole excitations around the Fermi zone. The resulting {\em dressed form factor} has now a macroscopic description in terms of the rapidities $\mu_{s_a}$ of the particles and holes separated from the Fermi boundaries.

The second step of our computation consists in summing up such contributions from different classes.
This means that we should sum up with respect to all possible values of the non-negative integer $n_0$, of the integer $\ell$, and also of the rapidities $\mu_{s_a}$ localized in a  small neighborhood $J_{\la_0}$ of the saddle point $\lambda_0$.
The latter sum can be replaced by an integral in the $L\to\infty$ limit according to the rule:
 \begin{equation}\label{sum-int}
   \frac{1}{L} \sum_{\mu_{s_a}\in J_{\la_0}} f(\mu_{s_a})\quad
   \mathop{\longrightarrow}_{L\rightarrow+\infty} \quad
   \Int_{J_{\la_0}} f(\mu) \, \rho(\mu) \, \dd \mu \qquad
   \text{for any regular function $f $},
\end{equation}
where $\rho(\mu)$ is the density \eqref{Lieb-eq}.
Since we have a multiple sum over the $n_0$ variables $\mu_{s_a}$, we obtain an
$n_0$-fold integral and, as we claimed before, the prefactor $L^{-n_0}$ in \eqref{Contr-class} does disappear:
\begin{multline}
  \lim_{L\to\infty}\sum_{\mu_{s_1}<\cdots<\mu_{s_{n_0}}\in J_{\la_0}}
  L^{-n_0} \,  \Delta^2(\{\mu_{s_a} \})\, e^{ix{\cal P}_{\mathrm{ex}}^{(0)}-it{\cal E}_{\mathrm{ex}}^{(0)}}
  \\
 =\frac1{n_0!}\,
   \Int_{J_{\la_0}}  \Delta^2(\{\mu\})\, e^{ix{\cal P}_{\mathrm{ex}}^{(0)}-it{\cal E}_{\mathrm{ex}}^{(0)}}
   \prod_{a=1}^{n_0}\rho(\mu_a)\, \dd\mu_a,
 \end{multline}
where ${\cal P}_{\mathrm{ex}}^{(0)}$ and ${\cal E}_{\mathrm{ex}}^{(0)}$ are given by \eqref{En-Mom1}.
Using standard saddle-point considerations, we expand the oscillating phase up to the second order term around $\la_0$:
\begin{equation}
   ix{\cal P}_{\mathrm{ex}}^{(0)}-it{\cal E}_{\mathrm{ex}}^{(0)}
   =ix(2\ell+\tau n_0) k_{{}_F}+ i \tau n_0(x\,p(\la_0)-t\,\varepsilon(\la_0))
     -it\tau \frac{u''(\la_0)}{2}\sum_{a=1}^{n_0}(\mu_{a}-\la_0)^2,
\label{u-saddle}
\end{equation}
and, using the fact that $u''(\lambda_0)>0$, we reduce our integral, in the $t\to\infty$, $x\to\infty$ limit, to the Gaudin--Mehta integral:
\begin{align}
 &\frac {1}{n_0!}\, \Int_{J_{\la_0}}
  \Delta^2(\{\mu\})\, \prod_{a=1}^{n_0}\rho(\mu_a)\,
  e^{-\frac{i t \tau u''(\la_0)}2(\mu_{a}-\la_0)^2}\,\dd\mu_a
                \nonumber\\
 &\hspace{2.6cm}
   =
  \frac{\rho^{n_0}(\lambda_0)\, e^{-i\tau \frac\pi 4 n_0^2\,\mathrm{sgn}(t)}}{n_0!} \,
  \Int_{\mathbb{R}}
  \Delta^2(\{\mu\})\,
  \prod_{a=1}^{n_0}e^{-\frac{|t|u''(\la_0)}2\mu_{a}^2} \, \dd\mu_a\cdot
  \left(1+O(t^{-1})\right)
                \nonumber\\
  &\hspace{2.6cm}
    =
  e^{-i\tau\frac\pi 4 n_0^2\,\mathrm{sgn}(t)} \,
  \frac{ (\sqrt{2\pi}\rho(\lambda_0))^{n_0} \, G(n_0+1)}{|tu''(\la_0)|^{n_0^2/2}}\cdot
  \left(1+O(t^{-1})\right).
 \label{Gaudin-Mehta}
\end{align}

Combining \eqref{Contr-class} with \eqref{Gaudin-Mehta} and taking the sum over $n_0$ and $\ell$,   we obtain the leading asymptotic terms for all  the oscillating harmonics of the two-point function at $x,t\to\infty$, $x/t=v(\lambda_0)$:
 \begin{multline}
  \moy{ {\cal O}^\dagger(x,t) \, {\cal O}(0,0) }
  = \sum_{n_0=0}^\infty\, \sum_{\ell=-\infty}^\infty
     \frac{(\sqrt{2\pi}\rho(\lambda_0))^{n_0} \, G(1+n_0)}{|t\, \varepsilon''(\la_0)-x\,p''(\la_0)|^{n_0^2/2}}
     \\
  \times
    e^{i\frac\pi 2\varphi_{\tau}( n_0,\ell)}
    \frac{  |\mathcal{F}^{\mathcal{O}}_{{}_{\tau n_0,\ell}} |^2 \,
               \exp\big[ix(2\ell+\tau n_0) k_{{}_F}+ i  \tau n_0(x\,p(\la_0)-t\,\varepsilon(\la_0))\big]}
           { | 2\pi  (x-v_{{}_F}t)|^{(F_{\tau n_0,\ell}^++\ell)^2}
             | 2\pi (x+v_{{}_F}t)|^{(F_{\tau n_0,\ell}^-+\ell+ \tau n_0)^2}  },
\label{Cor-fun-ell-pm-result}
 \end{multline}
where
\begin{equation}\label{phase-phi}
   \varphi_{\tau}( n_0,\ell)
   = \mathrm{sgn}(x-v_{{}_F}t)(F_{\tau n_0,\ell}^++\ell)^2
   -  \mathrm{sgn}(x+v_{{}_F}t)(F_{\tau n_0,\ell}^-+\ell+\tau n_0)^2
   -  \tau\,\mathrm{sgn}(t) \frac{n_0^2}{2}.
\end{equation}
We recall that $\tau=1$ in the space-like regime  and  $\tau=-1$ in the time-like regime, and that the values $F_{\tau n_0,\ell}^\pm$ of the shift function are given by \eqref{shift1}, \eqref{shift2}, where one should set $\mu_{s_a}=\lambda_0$.
Using the fact that  $|x/t|>v_{{}_F}$ in the space-like regime and that  $|x/t|<v_{{}_F}$ in the time-like regime, one can slightly simplify \eqref{phase-phi}:
\begin{align}
 &\varphi_{+}( n_0,\ell)
 =\mathrm{sgn}(x)\big[(F_{n_0,\ell}^++\ell)^2-(F_{n_0,\ell}^-+\ell+n_0)^2\big]-\mathrm{sgn}(t)\frac{n_0^2}{2},
         \label{phase-pl}\\
 &\varphi_{ -}(n_0,\ell)=- \mathrm{sgn}(t) \Big[(F_{-n_0,\ell}^++\ell)^2+(F_{-n_0,\ell}^-+\ell-n_0)^2 - \frac{n_0^2}{2}\Big] .
        \label{phase-mi}
 \end{align}

We keep in our result the sum over all possible values of $n_0$ and $\ell$ because of two reasons. First, each of these terms is associated with a different oscillating harmonic (and is leading in the corresponding harmonic). Second, generically it is not easy
to determine the leading term in the series \eqref{Cor-fun-ell-pm-result}. It may depend not only on the specific operator
$\mathcal{O}$ and  the position of the saddle point $\lambda_0$, but also on the original parameters of the model (the coupling
constant $c$ and the chemical potential $h$).

In the remaining part of this section, we specialize this general result to the case of the field-conjugated field and of the density-density dynamical correlation functions.

\subsection{Correlation functions of fields }

The asymptotic behavior of the correlation functions  $\moy{ \Psi(x,t)\,\Psi^\dagger (0,0) }$ and $\moy{ \Psi^\dagger(x,t) \, \Psi (0,0) }$ can be obtained directly from the general scheme presented in the last subsection.

The form factor sum for the two-point function $\moy{\Psi(x,t) \, \Psi^\dagger (0,0)}$ involves quasi-critical form factors with $\Delta N=1$.
Using \eqref{shift-ph}, \eqref{Z-Phi1}, \eqref{Z-Phi2}, we can rewrite the exponents of \eqref{Cor-fun-ell-pm-result} as
\begin{align}\label{ff-pl}
&F_{\tau n_0,\ell}^++\ell=\ell \mathcal{Z}+\frac 1{2\mathcal{Z}}+\tau n_0\,\Phi_+,\\
&F_{\tau n_0,\ell}^-+\ell+\tau n_0=\ell \mathcal{Z}- \frac 1{2\mathcal{Z}}+\tau n_0\,\Phi_-,
\label{ff-mi}
\end{align}
where $\mathcal{Z}$ is defined as the value of the dressed (fractional) charge $Z(\lambda)$ on the Fermi boundary:
\begin{equation}
   \mathcal{Z}=Z(q)=Z(-q).
\label{dcharge-bord}
\end{equation}
We have also introduced the following notations:
\begin{align}\label{Phi+}
  &\Phi_+ \equiv 
     \phi(q,-q)-\phi(q,\lambda_0)=-\phi(q,\lambda_0)+\frac{ \mathcal{Z}- \mathcal{Z}^{-1}}2,
     \\
  &\Phi_-  \equiv 
 1+\phi(-q,-q)-\phi(-q,\lambda_0) =-\phi(-q,\lambda_0)+\frac{ \mathcal{Z}+ \mathcal{Z}^{-1}}2,
  \label{Phi-}
\end{align}
where $\phi(\lambda,\mu)$ is the dressed phase \eqref{dphase}, and we used the identities (\ref{Z-Phi1}) and \eqref{Z-Phi2}

Now the general formula can be directly applied to give the following leading (in each harmonics) asymptotic behavior,
\begin{multline}
  \moy{\Psi(x,t)\,\Psi^\dagger (0,0) }
  = \sum_{n_0=0}^\infty\, \sum_{\ell=-\infty}^\infty
     \frac{(\sqrt{2\pi}\rho(\lambda_0))^{n_0} \, G(1+n_0)}{|t\, \varepsilon''(\la_0)-x\,p''(\la_0)|^{n_0^2/2}}
     \\
  \times
    e^{i\frac\pi 2\varphi_{\tau}( n_0,\ell)}
    \frac{  |\mathcal{F}^{\Psi^\dagger}_{{}_{\tau,\ell}} |^2 \,
               \exp\big[ix(2\ell+\tau n_0) k_{{}_F}+ i  \tau n_0(x\,p(\la_0)-t\,\varepsilon(\la_0))\big]}
           { | 2\pi  (x-v_{{}_F}t)|^{(\ell \mathcal{Z}+\frac 1{2\mathcal{Z}}+\tau n_0\,\Phi_+)^2}
             | 2\pi (x+v_{{}_F}t)|^{(\ell \mathcal{Z}- \frac 1{2\mathcal{Z}}+\tau n_0\,\Phi_-)^2}  }.
\label{Cor-Psi+}
 \end{multline}

To get some insight into this asymptotic behavior,  let us look at the possible values of the critical exponents $\delta_{n_0, \ell, \tau, \lambda_0}$  associated to the terms behaving as $  |x |^{-\delta_{n_0, \ell, \tau, \lambda_0}}$  in  \eqref{Cor-Psi+}. They are positive numbers given as the sum of three squares
\begin{equation}\label{delta}
\delta_{n_0, \ell, \tau, \lambda_0} =  \frac{n_0^2}{2} +   \Big(\ell \mathcal{Z}+\frac 1{2\mathcal{Z}}+\tau n_0 \Phi_+\Big)^2
            +\Big(\ell \mathcal{Z}-\frac 1{2\mathcal{Z}}+\tau n_0 \Phi_-\Big)^2.
\end{equation}
For $n_0=0$, the critical  exponents do not depend on $\tau$, $\lambda_0$ and we get
\begin{equation}\label{delta0}
\delta_{0, \ell} =  2 \ell^2 \mathcal{Z}^2 + \frac{1}{2 \mathcal{Z}^2}.
\end{equation}

Hence, the $n_0=0$ series over $\ell$ in our  result  \eqref{Cor-Psi+} fully reproduces the Luttinger liquid theory and CFT predictions and provides in addition the exact values of the associated amplitudes. In this case, the dominant term is given obviously by $\ell=0$ with critical exponent $1/2 \mathcal{Z}^2 \le 1/2$ as in the QNLS model with finite positive coupling $\mathcal{Z} \ge 1$ ($\mathcal{Z}=1$
only in the free fermion limit $c\to+\infty$).

The  rest of the series, namely the terms with $n_0\ge 1$,  corresponds to the contribution of the saddle point  located in the non-linear part of the spectrum (separated from the Fermi boundary). It is important to stress that CFT and Luttinger liquid theory fail to predict such contributions and the corresponding oscillatory harmonics which explicitly depend on the saddle point $\lambda_0$ (hidden in $\tau \Phi_{\pm}$).

To see the possible relevance of these additional terms, let us discuss now the relative values of the different critical exponents.  For  $n_0\ge 1$, $\delta_{n_0, \ell, \tau, \lambda_0}  \ge 1/2$, whereas $1/2 \mathcal{Z}^2 < 1/2$ (it is equal to $1/2$ only at free fermion point). Hence, the term $n_0=0$, $\ell=0$  gives the dominant contribution to the asymptotic series  \eqref{Cor-Psi+},  which confirms the CFT prediction.

The origin of  the first subleading term of the series   \eqref{Cor-Psi+} is more involved. In the CFT and Luttinger liquid theory it  is given by $n_0=0$, $\ell=\pm 1$ and the corresponding critical exponent  $2 \mathcal{Z}^2 + 1/{2 \mathcal{Z}^2}$ is  greater than or equal to $5/2$. It so happens that the leading saddle point contribution can be dominant with respect to this first subleading CFT contribution  (see also \cite{Koz11}).

To see this, let us consider a very simple example, namely the free fermion point 
where $\mathcal{Z}=\Phi_-=1$ and $\Phi_+=0$. In this case, all the dependence of the critical exponents  $\delta_{n_0, \ell, \tau, \lambda_0}$ on $\lambda_0$  reduces to the value of $\tau=\pm$, namely it just indicates whether we are in the space-like ($\tau=1$) or in the time-like ($\tau=-1$) asymptotic regimes. We have for the free fermion point critical exponents,
\begin{equation}\label{delta-ff}
\delta_{n_0, \ell, \tau} =  \frac{n_0^2}{2} +   \Big(\ell +\frac 1{2} \Big)^2
            +\Big(\ell -\frac 1{2}+\tau n_0 \Big)^2.
\end{equation}
For a given $n_0$, $\delta_{n_0, \ell, \tau}$  as a function of $\ell$ (considered as a real variable) would reach its minimal value at $\ell=-\tau n_0/2$. Hence for $n_0=1$, the dominant term is realized with the nearest integer values $\ell=0, -\tau$ with corresponding critical exponents $\delta_{1, 0, \tau}$ and $\delta_{1, -\tau, \tau}$.

In the space-like regime, $\tau=1$, and the leading critical exponents coming from the saddle point are  $\delta_{1, 0, 1}=1$ and $ \delta_{1, -1, 1}=1$, which produce two dominant terms in \eqref{Cor-Psi+} with respect to the first subleading CFT   critical exponents $\delta_{0, \pm1, 1}=5/2$. 
Hence in this  case, the saddle point provides the first oscillatory  subleading contribution to the asymptotic behavior of the series \eqref{Cor-Psi+}.

In the time-like regime, $\tau=-1$, and the leading critical exponents coming from the saddle point  are  $\delta_{1, 0, -1}=3$ and $\delta_{1, 1, -1}=3$. Hence, they are subdominant with respect to $\delta_{0, \pm1, -1}=5/2$.

 The two-point function $\moy{ \Psi^\dagger(x,t)\,\Psi (0,0)}$ can be treated in a very similar way. The relevant form factors are those with $\Delta N =-1$. The corresponding exponents in this case are
\begin{align}\label{ff-pl1}
  &F_{\tau n_0,\ell}^++\ell=\ell \mathcal{Z}-\frac 1{2\mathcal{Z}}+\tau n_0\,\Phi_+,\\
  &F_{\tau n_0,\ell}^-+\ell+\tau n_0=\ell \mathcal{Z}+ \frac 1{2\mathcal{Z}}+\tau n_0\,\Phi_-,
\label{ff-mi1}
\end{align}
leading to the following asymptotic expansion of the correlation function:
\begin{multline}
  \moy{ \Psi^\dagger(x,t)\, \Psi (0,0)}
  = \sum_{n_0=0}^\infty\, \sum_{\ell=-\infty}^\infty
     \frac{(\sqrt{2\pi}\rho(\lambda_0))^{n_0} \, G(1+n_0)}{|t\, \varepsilon''(\la_0)-x\,p''(\la_0)|^{n_0^2/2}}
     \\
  \times
    e^{i\frac\pi 2\varphi_{\tau}( n_0,\ell)}
    \frac{  |\mathcal{F}^{\Psi}_{{}_{\tau,\ell}} |^2 \,
               \exp\big[ix(2\ell+\tau n_0) k_{{}_F}+ i  \tau n_0(x\,p(\la_0)-t\,\varepsilon(\la_0))\big]}
           { | 2\pi  (x-v_{{}_F}t)|^{(\ell \mathcal{Z}-\frac 1{2\mathcal{Z}}+\tau n_0\,\Phi_+)^2}
             | 2\pi (x+v_{{}_F}t)|^{(\ell \mathcal{Z}+\frac 1{2\mathcal{Z}}+\tau n_0\,\Phi_-)^2}  },
\label{Cor-Psi-}
 \end{multline}
the phase $\varphi_{\tau}(n_0,\ell)$ being now built out of the shift functions
\eqref{ff-pl1}, \eqref{ff-mi1}.

The analysis of this asymptotic series is similar to the one given above for \eqref{Cor-Psi+}. The CFT and Luttinger liquid predictions are reproduced for $n_0=0$ and the term with $n_0=0$ and $\ell=0$ is dominant with critical exponent $1/2 \mathcal{Z}^2 \le 1/2$. As for the first subleading term, the discussion is very similar to the previous case with one notable difference: taking the example of the free fermion point, the first subleading oscillatory contribution is provided by the saddle point, with $n_0=1$,  in the time-like regime, while in the space-like regime it comes from the terms $n_0=0$ and $\ell=\pm 1$.

\subsection{Correlation function of densities\label{subsec-corr-dens}}

The asymptotic behavior of the time-dependent  density-density correlation function can also be inferred from the general framework developed in Section~\ref{sec-general}.
There is however some peculiarity in this case due to the fact that  $F^{\pm}_{\tau n_0,\ell}=0$ for $\ell=n_0=0$.
As a consequence, the method described above, when applied directly to the two-point function, only enables one to obtain the trivial, constant part $D^2$ of the non-oscillating contribution to the asymptotic expansion (the oscillating terms do not suffer from such effects).
This apparent problem can be circumvented by considering, just as in the time-independent case, a small modification of the initial form factor series.

Namely, as in \cite{KitKMST11b} (see also \cite{KozT11}), it is convenient to introduce the generating function
$\moy{ e^{itH}e^{2\pi i\alpha{\cal Q}(x)}e^{-itH_\alpha} }$, where $\alpha$ is a real number (twist).
Here we define the operator $\mathcal{Q}(x)$ as
\begin{equation}\label{def-Q}
    {\cal Q}(x)=\int\limits_{0}^x j(y)\, \dd y,
\end{equation}
and $H_\alpha$ corresponds to the quantum non-linear Schr\"odinger Hamiltonian \eqref{Ham} for which we impose twisted quasi-periodic boundary conditions  $\Psi(x+L)=e^{2\pi i\alpha}\Psi(x)$ instead of periodic ones.
The time dependent correlation function of densities can be obtained from the  derivatives of this generating function (see Appendix~\ref{twist}):
\begin{equation}
  \moy{ j(x,t) \, j(0,0) }
   =-\frac 1{8\pi^2}\left.\frac{\partial^2}{\partial\alpha^2}\frac{\partial^2}{\partial x^2}
       \moy{ e^{itH}e^{2\pi i\alpha{\cal Q}(x)}e^{-itH_\alpha} }\right|_{\alpha=0}\;.
\label{gen_f_deriv}
\end{equation}

The form factor approach described in Subsection~\ref{sec-general} can be very simply adapted to the evaluation of this generating function by introducing the eigenstates $\ket{\psi'_{\alpha}(\{\widehat\mu\})}$ of the Hamiltonian $H_\alpha$ with quasi-periodic boundary conditions.
These eigenstates are given by the solutions of the twisted Bethe equations:
\begin{equation}\label{eq-Bethe-alpha}
 L \, p_0(\widehat\mu_{\ell_j})+\sum_{k=1}^{N}\theta(\widehat\mu_{\ell_j}-\widehat\mu_{\ell_k})
 =2\pi\Big(\ell_j+\alpha-\frac{N+1}{2}\Big),
 \qquad j=1,\ldots,N'.
\end{equation}
One can show (see Appendix~\ref{twist}) that the generating function has the following expansion over
the complete set of twisted eigenstates $\ket{\psi'_{\alpha}(\{\widehat\mu\})}$:
 \begin{equation}\label{Expan-GF}
    \moy{ e^{itH}e^{2\pi i\alpha{\cal Q}(x)}e^{-itH_\alpha} }
    =
    \sum_{ \ket{\psi'_\alpha(\{\mu\}) } }
    e^{-i t{\mathcal{E}}_{\mathrm{ex}} +ix {\mathcal{P}}_{\mathrm{ex}}  } \,
    \bigl( \mathcal{F}^\alpha(\{\widehat\mu\})\bigr)^2,
\end{equation}
where
 \begin{equation}\label{def-F}
    \mathcal{F}^\alpha(\{\widehat\mu\})
    =\frac{ \moy{\psi'_\alpha(\{\widehat\mu\}) \, | \, \psi_g} }{{\|\psi'_\alpha(\{\widehat\mu\})\| \, \|\psi_g\|}}
    =\frac{ \moy{\psi_g \, | \, \psi'_\alpha(\{\widehat\mu\}) } } {{\|\psi'_\alpha(\{\widehat\mu\})\| \, \|\psi_g\|}},
 \end{equation}
and  ${\mathcal{E}}_{\mathrm{ex}} $ and $ {\mathcal{P}}_{\mathrm{ex}}$ are defined as in \eqref{Mom-Ener-finite} in terms of the rapidities $\widehat{\mu}_{ \ell_a}$ solution to \eqref{eq-Bethe-alpha}.

The general scheme of Section~\ref{sec-ff-exp} applies to the computation of the asymptotic expansion in the $\alpha$-twisted case with only slight modifications.
The shift functions  $F_{\tau n_0,\ell}^\pm$ can be directly obtained from the Bethe equations.
The corresponding exponents  are
\begin{align}
 &F_{\tau n_0,\ell}^++\ell=(\alpha+\ell) \mathcal{Z}+\tau n_0\,\Phi_+,\\
 &F_{\tau n_0,\ell}^-+\ell+\tau n_0=(\alpha+\ell) \mathcal{Z}+\tau n_0\,\Phi_-,
\end{align}
and the  phases $\varphi_{\tau}(n_0,\ell)$ take the following form according to whether we are in the space-like or time-like regime:
\begin{align}\label{phase-plj}
  &\varphi_{ +}(n_0,\ell)
    =\mathrm{sgn}(x)[((\ell+\alpha)\mathcal{Z}+n_0\tau\Phi_+)^2
     - ((\ell+\alpha)\mathcal{Z}+n_0\tau\Phi_-)^2]-\mathrm{sgn}(t) \frac{n_0^2}{2} ,
 \\
  &\varphi_{-}(n_0,\ell)
    =-\mathrm{sgn}(t)\Big[((\ell+\alpha)\mathcal{Z}+n_0\tau\Phi_+)^2
      + ((\ell+\alpha)\mathcal{Z}+n_0\tau\Phi_-)^2-\frac{n_0^2}{2}\Big] .
      \label{phase-mij}
 \end{align}
It leads to the following asymptotic behavior for the generating function:
\begin{multline}
  \moy{ e^{itH}e^{2\pi i\alpha{\cal Q}(x)} e^{-itH_\alpha} }
  =  \sum_{n_0=0}^\infty\,
      \sum_{\ell=-\infty}^\infty
      \frac{(\sqrt{2\pi}\rho(\lambda_0))^{n_0} \, G(1+n_0)}{|t\, \varepsilon''(\la_0)-x\,p''(\la_0)|^{n_0^2/2}}
     \\
     \times
     \frac{ \big|\mathcal{F}^{\alpha}_{{}_{\tau n_0,\ell}}\big|^2
               e^{ix(2\ell +2\alpha+n_0\tau) k_{{}_F}+ i n_0\tau\,(x\,p(\la_0)-t\,\varepsilon(\la_0))
              +i\frac\pi2 \varphi_{\tau}(n_0,\ell)}}
             { \big( 2\pi |x-v_{{}_F}t|\big)^{\left((\ell+\alpha) \mathcal{Z}+n_0\tau \Phi_+\right)^2}
               \big(2\pi |x+v_{{}_F}t|\big)^{\left((\ell+\alpha) \mathcal{Z}+n_0\tau \Phi_-\right)^2}  }.
\label{Cor-fun-ell-pm-result-alpha}
\end{multline}
Here $\mathcal{F}^{\alpha}_{{}_{\tau n_0,\ell}}$ is the thermodynamic  limit of the properly normalized (\textit{cf}   \eqref{basic_ff}) $\alpha$-twisted
scalar product \eqref{def-F} for the specific solution of \eqref{eq-Bethe-alpha} which corresponds to a choice of the integers $\ell_j$ as described on page~\pageref{simplest-state}.

It then remains to differentiate \eqref{Cor-fun-ell-pm-result-alpha} with respect to $\alpha$ and $x$ so as to get the asymptotic behavior of the  two-point function.
Just as in  the time-independent case, one should consider the case $\ell=n_0=0$ separately.
Indeed, $|\mathcal{F}^{\alpha}_{{}_{\tau n_0,\ell}}|^2$
has a second order zero at $\alpha=0$ except in the case  $\ell=n_0=0$ for which
$|\mathcal{F}^{\alpha}_{{}_{0,0}}|^2=1$ at $\alpha=0$  (see \cite{KitKMST11b} for details).
Using \eqref{Fa-densityFF} we obtain
\begin{multline}
\moy{ j(x,t) \, j(0,0)}
  =  \left(\frac{k_{{}_F}}{\pi}\right)^2
        - \frac{\mathcal{Z}^2}{2\pi^2}\frac{x^2+v^2_{{}_F}t^2}{ (x^2-v^2_{{}_F}t^2 )^2}
        +\sul_{n_0=0}^\infty\,
          \sum_{\ell=-\infty\atop{|\ell|+n_0>0}}^\infty
 \frac{(\sqrt{2\pi}\rho(\lambda_0))^{n_0} \, G(1+n_0)}{|t\, \varepsilon''(\la_0)-x\,p''(\la_0)|^{n_0^2/2}}
         \\
        \times
        \frac{ \big|\mathcal{F}^{j}_{{}_{\tau n_0,\ell}}\big|^2
                  e^{ix(2\ell +n_0\tau) k_{{}_F}+ i \tau n_0 (x\,p(\la_0)-t\,\varepsilon(\la_0))
                        +i\frac\pi2 \varphi_\tau(n_0,\ell)}}
                { \big( 2\pi |x-v_{{}_F}t|\big)^{\left(\ell \mathcal{Z}+\tau n_0 \Phi_+\right)^2}
                  \big(2\pi |x+v_{{}_F}t|\big)^{\left(\ell \mathcal{Z}+\tau n_0 \Phi_-\right)^2}  } \; .
\label{currents}
\end{multline}
We stress that, on the level of \eqref{currents} one should set $\alpha=0$ in the expressions \eqref{phase-plj}, \eqref{phase-mij} for $\varphi_\tau(n_0,\ell)$.

Let us look now at the possible values of the critical exponents $\delta_{n_0, \ell, \tau, \lambda_0}$  associated to the terms behaving as $  |x |^{-\delta_{n_0, \ell, \tau, \lambda_0}}$  in  \eqref{currents}. Besides the constant term, the non-oscillating term has critical exponent equal to $2$. For  oscillating terms in \eqref{currents}  the critical exponents are given as the sum of three squares
\begin{equation}\label{deltacurrent}
\delta_{n_0, \ell, \tau, \lambda_0} =  \frac{n_0^2}{2} +   \Big(\ell \mathcal{Z}+\tau n_0 \Phi_+\Big)^2
            +\Big(\ell \mathcal{Z}+\tau n_0 \Phi_-\Big)^2.
\end{equation}
For $n_0=0$,  they do not depend on $\tau$, $\lambda_0$ and we get
\begin{equation}\label{delta0currrent}
\delta_{0, \ell} =  2 \ell^2 \mathcal{Z}^2.
\end{equation}

Hence, as for the correlation functions of the  fields, the series over $\ell$ obtained for $n_0=0$ reproduces the CFT and Luttinger liquid predictions, with the exact values of the amplitudes given here in terms of specific (renormalized) form factors of the 
density. The dominant term of the asymptotic expansion \eqref{currents} is evidently the constant term. The origin of the first subleading term is more involved in general.

The terms with $n_0>1$ are subdominant with respect to the non-oscillating contribution which has critical exponent equal to $2$. Hence, after the constant term,
the leading behavior is produced either by the $n_0=0$ and $n_0=1$ terms or by the non-oscillating term.
For $n_0=0$, the leading terms correspond to  $\ell=\pm 1$ with critical exponent $2 \mathcal{Z}^2 \ge 2$, hence greater than or equal to the critical exponent of the non-oscillating term,  the terms with $n_0=0$ and $\ell>1$ being evidently subleading.
It is more complicated to select the leading contribution for $n_0=1$ and to compare it in general to $2$ or  $2 \mathcal{Z}^2$ since it depends on the regime and on the values of the constants $\Phi_\pm$.

To see at least the possible relevance of these different terms, let us consider again the example of the free fermion point. There, the leading non-constant terms correspond to $n_0=1$, $\ell=0$ and $n_0=1$, $\ell=-\tau$, both with the critical exponent $3/2$, which is dominant with respect to the non-oscillating term and  to the first subleading oscillating term predicted from CFT (both with critical exponent $2$). In this example, the saddle point provides the main contribution after the constant term a soon as $x/t$ takes a finite value in the asymptotic regime. Hence, in this case,  Luttinger liquid theory and CFT fail to predict the leading asymptotic behavior of the density correlation function (see also \cite{KozT11}).

\section{Behavior of dynamical response functions on the particle/hole excitation thresholds}
\label{sec-edge}

In this section, we use the form factor expansion  to study  the dynamical zero-temperature density structure factor $S(k,\om)$ (DSF) and spectral function $A(k,\om)$, which are defined in terms of the space and time Fourier transform of the two-point dynamical correlation functions:
\begin{align}
  &S(k,\om)=\Int_{\Rset^2} e^{i(\om t-kx)}\, \moy{j(x,t)\, j(0,0) }\, \dd x\, \dd t, \label{DSF}\\
  &A(k,\om)=\frac{1}{\pi}\, \mathrm{sign}(\om)\;
  \Re \!  \Int_{\Rset^2}   e^{i(\om t-kx)}\,
   \moy{T\,\Psi(x,t)\,\Psi^\dagger(0,0)  } \, \dd x\, \dd t. \label{spectr-fct}
\end{align}
Here $T$ denotes the time-ordering operator.
$S(k,\om)$ describes the probability to excite the ground state with momentum and energy transfer $(k,\om)$, whereas
$A(k,\om)$ describes the response of the system to the addition of a physical particle or hole with momentum $k$ and energy $\om$.
Both quantities can be measured experimentally \cite{StaCGIGPK99,Dua06,DaoGDSC07}.

\subsection{General scheme}
\label{subsec-TF}

Let us consider the space and time Fourier transform of a two-point function  of the type \eqref{def-2pt}.
When evaluating this quantity at some point $(k,\om)$ from its form factor series, we see that the only contributions to the form factor sum come from states $\ket{\psi'}$ with excitation momentum and energy precisely given by $(k,\om)$:
\begin{equation}\label{loc-TF}
   \Int_{\Rset^2} \dd x\, \dd t \, e^{i(\om t-kx)} \moy{\mc{O}^{\dagger}(x,t)\,\mc{O}(0,0) }
   =\sum_{|\psi'\rangle} \delta(\om-\mc{E}_{\text{ex}})\,\delta(k-\mc{P}_{\text{ex}})
    \left| \frac{ 
            \bra{\psi'}\,\mc{O}(0,0)\,\ket{\psi_g}}
           { || \psi' || \; ||\psi_g|| } \right|^2 .
\end{equation}
This is due to the fact that, in the form factor sum \eqref{ff-series}, the only $x$ and $t$ dependence is contained in the phase factor.
Note in particular that the quantity \eqref{loc-TF} vanishes for $\om<0$, because\footnote{Note however that, due to the presence of the time-ordering operator, the spectral function \eqref{spectr-fct} does \textit{not} vanish for $\om<0$.}  $\mc{E}_{\text{ex}}\ge 0$.  It is also invariant under the change $k\to -k$, since the existence of a state
with $\mc{P}_{\text{ex}}$ implies the existence of a state with $-\mc{P}_{\text{ex}}$. This means that we can restrict our study to the domain $k>0,\om>0$.

We shall consider the behavior of the above quantity in the vicinities of Lieb's one-particle or one-hole excitation spectra \cite{Lie63}.
In other words,  let $\la$ denote the rapidity of an arbitrary particle ($\la>q$) or an arbitrary hole  ($\la\in ]-q,q[$) located at finite distance from the endpoints $\pm q$ of the Fermi zone.
Let $k_p=p(\la)-k_{{}_F}$ (resp. $k_h=k_{{}_F}-p(\la)$) be the momentum of the excitation corresponding to a particle at $\la$ and a hole at $q$ (resp. to a hole at $\la$ and a particle at $q$), the corresponding energy being equal to $\veps_p=\veps(\la)$ (resp. to $\veps_h=-\veps(\la)$).
We are interested in the $\delta\om\to 0$ behavior of \eqref{loc-TF} when
\begin{equation}\label{kom-part}
  k= k_p=p(\la)-k_{{}_F} \quad \text{and} \quad \om=\veps_p+\delta\om=\veps(\la)+\delta\om
   \qquad\quad (\text{for}\quad \la>q),
\end{equation}
i.e. in the vicinities of the one-particle excitation threshold, or when
\begin{equation}\label{kom-hole}
  k=k_h=k_{{}_F}-p(\la) \quad \text{and} \quad \om=\veps_h+\delta\om=-\veps(\la)+\delta\om
   \qquad (\text{for}\quad \la\in]-q,q[),
\end{equation}
i.e. in the vicinities of the one-hole excitation threshold.

Recall that we assume the sound velocity   $v(\lambda)$ to be a strictly monotonic function: $v'(\lambda)>0$. Then one
can easily see \cite{CorDZ09} that a given combination of momentum and energy  $(k_p,\veps_p)$ (resp. $(k_h,\veps_h)$)
belonging to the threshold can only be realized by a unique particle/hole configuration at the macroscopic level:  {\em one} particle (resp. hole) with rapidity $\la$,  {\em one} hole (resp. particle) on the right Fermi boundary $+q$. On the microscopic level, however, one
can dress this single state by an arbitrary number of additional particle-hole excitations with rapidities accumulating on the two endpoints $\pm q$ of the Fermi zone, so that the total momentum and total energy of these additional particle-hole excitations is effectively zero at the thermodynamic limit. It means that we should consider quasi-critical states of the $\mathbf{P}_{\pm 1,\mp 1}$ class, with only one rapidity $\mu$ separated from the Fermi boundaries. If we admit a small deviation of the energy
$\veps_p+\delta\om$ (resp. $\veps_h+\delta\om$), then  $\mu$  belongs to a small neighborhood $J_\la$ of $\la$, tending to $\la$ in the $\delta\om\to 0$ limit.
Hence, we should sum up the contributions of the corresponding
$\mathbf{P}_{\pm 1,\mp 1}$ classes, for all such values of $\mu$.

In fact, all this was already done in Section~\ref{sec-asympt} for a more general case. The only difference is that now the ratio $x/t$ is not fixed and the point $\lambda$ is no longer a saddle point. Therefore, to take into account the variation of the oscillating phase in the small neighborhood $J_\la$ of $\la$, it is no longer necessary to develop the function $u(\mu)$ up to the second order as in \eqref{u-saddle}. Instead we simply linearize it around $\la$:
\begin{equation}
  x p(\mu)-t\veps(\mu)=x p(\la)-t\veps(\la)+[xp'(\la)-t\veps'(\la)] \, (\mu-\la) +\dots.
\end{equation}
In all other respects, one can compute the contribution of the $\mathbf{P}_{\pm 1,\mp 1}$ classes form factors by the
method described in Section~\ref{sec-asympt}.

For example, the contribution of the form factor sum in the vicinities of the particle threshold (that is the contribution of the sum over the corresponding $\mathbf{P}_{1,-1}$ classes) is
\begin{multline}
 \moy{ \mathcal{O}^\dagger(x,t)\,\mathcal{O}(0,0) }_{k_p,\varepsilon_p+\delta\omega}
 =\lim_{L\to\infty}
    \sum_{\mu\in J_\la} L^{-1}
        e^{i x\,  p'(\la)\, (\mu -\la )-it \veps'(\la)\,  (\mu -\la ) }
   \\
   \times
    \frac{ L^{-\theta_{1,-1}+1}\,
              \big|\mathcal{F}^{\mathcal{O}}_{{}_{1,-1}}\big|^2\,
             e^{ -ix k_{{}_F}+i x\,p(\la)-it\,\varepsilon(\la)} }
           { \big[1-e^{ \frac{2\pi i }{L}(x-v_{{}_F}t)} \big]^{(F_{1,-1}^+ -1)^2} \,
             \big[1-e^{- \frac{2\pi i }{L}(x+v_{{}_F}t)}\big]^{(F_{1,-1}^-)^2}  }.
\label{contr-la}
\end{multline}
Recall that, as discussed previously, one should use the regularization $t\to t-i0$ when applying the summation identity \eqref{magic-formula}.
Taking the thermodynamic limit in \eqref{contr-la}, we obtain
\begin{multline}
 \moy{ \mathcal{O}^\dagger(x,t)\,\mathcal{O}(0,0) }_{k_p,\varepsilon_p+\delta\omega}
 =
   \rho(\la)\,
   \Int_{J_\la} \dd\mu \,
        e^{i x p'(\la)\, (\mu -\la )-it \veps'(\la)\, (\mu -\la ) }
   \\
   \times
       \frac{
              \big|\mathcal{F}^{\mathcal{O}}_{{}_{1,-1}}\big|^2  \;
             e^{ -ix k_{{}_F}+i x p(\la)-it \varepsilon(\la)} }
           { [-2\pi i (x-v_{{}_F}t)] ^{(F_{1,-1}^+ -1)^2} \,
              [ 2\pi i (x+v_{{}_F}t)]^{(F_{1,-1}^-)^2}  } ,
\label{contr-part}
\end{multline}
where we have used \eqref{sum-int} and replaced $\rho(\mu)$ by its value $\rho(\lambda)$ at the localization point.
In contrast to  \eqref{2-line}, we prefer to present the result in terms of $x\pm v_{{}_F}t$ instead of their absolute values $|x\pm v_{{}_F}t|$, keeping in mind that $t$ is slightly shifted to the lower half-plane. As we will see very soon,
such form is more convenient for taking the Fourier transform.

The space and time Fourier transform of the above correlation function  in the vicinity of the one-particle excitation threshold, i.e. at the point $(k,\om)\equiv (k_p,\veps_p+\delta\om)$ (see \eqref{kom-part}), is therefore given by the following quantity:
\begin{multline}\label{TF-part1}
   \Int_{\Rset^2} \dd x\, \dd t \, e^{i(\om t-kx)}
     \moy{ \mathcal{O}^\dagger(x,t)\,\mathcal{O}(0,0) }_{k_p,\varepsilon_p+\delta\omega}
 = \rho(\la)\, \big|\mathcal{F}^{\mathcal{O}}_{{}_{1,-1}}\big|^2
     \\
     \times
    \Int_{-\eps_{\delta\om}}^{\eps_{\delta\om}} \dd\nu \,  \Int_{\Rset^2} \dd x\, \dd t \,
      \frac{e^{i x \nu p'(\la)+it [\delta\om- \nu \veps'(\la)] } }
           { [-2\pi i (x-v_{{}_F}t)] ^{(F_{1,-1}^+ -1)^2}\,
              [ 2\pi i (x+v_{{}_F}t)]^{(F_{1,-1}^-)^2}   },
\end{multline}
in which we have set $J_\la=[\la-\eps_{\delta\om},\la+\eps_{\delta\om}]$, for some $\eps_{\delta\om}>0$ such that both $\eps_{\delta\om}$ and $\delta\om/\eps_{\delta\om}$ tend to zero in the limit $\delta\om\to 0$.

Similarly, in the vicinity of the one-hole excitation threshold, i.e. for  $(k,\om)\equiv (k_h,\veps_h+\delta\om)$ given by \eqref{kom-hole}, we get
\begin{multline}\label{TF-hole1}
   \Int_{\Rset^2} \dd x\, \dd t \, e^{i(\om t-kx)}
     \moy{ \mathcal{O}^\dagger(x,t)\,\mathcal{O}(0,0) }_{k_h,\varepsilon_h+\delta\omega}
 = \rho(\la)\, \big|\mathcal{F}^{\mathcal{O}}_{{}_{-1,1}}\big|^2
     \\
     \times
    \Int_{-\eps_{\delta\om}}^{\eps_{\delta\om}} \dd\nu \,  \Int_{\Rset^2} \dd x\, \dd t \,
      \frac{e^{-i x \nu p'(\la)+it [\delta\om+\nu \veps'(\la)] } }
           { [-2\pi i (x-v_{{}_F}t)] ^{(F_{-1,1}^++1)^2}\,
              [ 2\pi i (x+v_{{}_F}t)]^{(F_{-1,1}^-)^2}   }.
\end{multline}

Integrals over $x$ and $t$ in \eqref{TF-part1} and \eqref{TF-hole1} are therefore of the type
\begin{align}
   I_{\alpha_+,\alpha_-}(E,P)
    & \equiv \Int_{\Rset^2} \dd x\, \dd t \,
      \frac{e^{-i P x+i E t  } }
           { [- 2\pi i (x-v_{{}_F}t+i0)] ^{\alpha_+}\,
              [   2\pi i (x+v_{{}_F}t-i0)]^{\alpha_-}   }
             \label{int-EP} \\
    & = H(E+v_{{}_F} P) \, H(E-v_{{}_F} P) \,
        \frac{2\pi\, [E +v_{{}_F} P]^{\alpha_+ -1}\,    [E -v_{{}_F} P]^{\alpha_- -1} }
                {[4\pi   v_{{}_F}]^{\alpha_+ +\alpha_- -1}\, \Gamma(\alpha_+)\,\Gamma(\alpha_-)},
                \label{result-int}
\end{align}
where $H$ denotes the Heaviside step function.
The expression \eqref{result-int} for \eqref{int-EP} can easily be obtained through the change of variables
$(\varphi,\psi)=(x-v_{{}_F}t,x+v_{{}_F}t)$ which separates the two-fold integral \eqref{int-EP} into a product of two  integrals:
\begin{equation}\label{int-TF2}
   I_{\alpha_+,\alpha_-}(E,P)=\frac{1}{2 v_{{}_F}}\,
        I_{\alpha_+}\Bigl(\frac{E+v_{{}_F}P}{2v_{{}_F}}\Bigr) \cdot
        I_{\alpha_-}\Bigl(\frac{E-v_{{}_F}P}{2v_{{}_F}}\Bigr),
\end{equation}
with
\begin{equation}\label{int2}
   I_{\alpha}(b)=\Int_{\Rset}\dd\psi\frac{e^{-i b \psi}}{[-2\pi i(\psi+i0)]^\alpha}
                                  =\Int_{\Rset}\dd\psi\frac{e^{i b \psi}}{[2\pi i(\psi-i0)]^\alpha}.
\end{equation}
The latter can be easily calculated via substitution
\begin{equation}\label{int2a}
   [\mp i(\psi\pm i0)]^{-\alpha}=\frac{1}{\Gamma(\alpha)}\int_0^\infty e^{\pm i\psi s}s^{\alpha-1}\,\dd s,
   \end{equation}
leading to
\begin{equation}\label{int3}
   I_{\alpha}(b)= H( b)\,\frac{(2\pi)^{1-\alpha} \, b^{\alpha-1}}{\Gamma(\alpha)}.
\end{equation}
%

\subsubsection{Behavior of the Fourier transform of the two-point function close to the particle excitation threshold}

To obtain the behavior of the Fourier transform in the vicinity of the one-particle excitation threshold, we substitute in \eqref{result-int}:  $\alpha_+=(F_{1,-1}^+-1)^2$, $\alpha_-=(F_{1,-1}^-)^2$, and $E=\delta\om-\nu\veps'(\la)$, $P=-\nu p'(\la)$. Hence we have
\begin{equation}
  E \pm v_{_F}P=\delta\om-\nu\veps'(\la)\mp v_{_F} \nu p'(\la)
                           = \delta\om - \nu p'(\la) [ v\pm v_{_F} ] \; .
\end{equation}
Above, we have used the expression \eqref{v-sound} for the sound velocity $v=v(\lambda)$.
The two Heaviside functions restrict the domain of integration on $\nu$ such that
\begin{equation}\label{rest-nu}
   \nu<\frac{\delta\om}{p'(\la)[v\pm v_{_F}]},
   \qquad \text{i.e.}\qquad
   \begin{cases}
        {\displaystyle \nu<\frac{\delta\om}{p'(\la)[v + v_{_F}]}} & \text{if $\delta\om>0$,}
        \smallskip\\
        {\displaystyle \nu<\frac{\delta\om}{p'(\la)[v - v_{_F}]} } & \text{if $\delta\om<0$.}
   \end{cases}
\end{equation}
In \eqref{rest-nu}, we have used the fact that $v>v_{_F}$ (recall that $\la>q$).
Hence,
\begin{align}\label{TF-part2}
   \Int_{\Rset^2} \dd x\, \dd t \, e^{i(\om t-kx)}
    & \moy{ \mathcal{O}^\dagger(x,t)\,\mathcal{O}(0,0) }_{k_p,\varepsilon_p+\delta\omega}
       =  \frac{2\pi\, \rho(\la) \, \big|\mathcal{F}^{\mathcal{O}}_{{}_{1,-1}}\big|^2 }
            {[4\pi v_{_F}]^{\alpha_+ +\alpha_- -1}\,\Gamma(\alpha_+)\,\Gamma(\alpha_-)} \nonumber\\
    &\qquad\times \Bigg\{
     H(\delta\om) \int_{-\eps_{\delta\om}}^{\frac{\delta\om}{p'(\la)[v+v_{_F}]}}
     + H(-\delta\om) \int_{-\eps_{\delta\om}}^{\frac{\delta\om}{p'(\la)[v-v_{_F}]}  }
     \Bigg\}\nonumber\\
        &\qquad\times [\delta\om - \nu p'(\la) ( v + v_{_F} )]^{\alpha_+ -1}\,
        [ \delta\om - \nu p'(\la) ( v- v_{_F} )]^{\alpha_- -1} \,\dd\nu.
\end{align}
Changing  variables $w=(v-v_{_F})[ \delta\om - \nu p'(\la) ( v + v_{_F} )]/(2v_{_F}\delta\om)$ in the first integral, and $w=-(v+v_{_F})[ \delta\om - \nu p'(\la) ( v - v_{_F} )]/(2v_{_F}\delta\om)$ in the second integral, we obtain
\begin{multline}\label{TF-part3}
   \Int_{\Rset^2} \dd x\, \dd t \, e^{i(\om t-kx)}
     \moy{ \mathcal{O}^\dagger(x,t)\,\mathcal{O}(0,0) }_{k_p,\varepsilon_p+\delta\omega}
 \sim \frac{\,  \big|\mathcal{F}^{\mathcal{O}}_{{}_{1,-1}}\big|^2\,
 \left|\frac{\delta\om}{2\pi}\right|^{\alpha_++\alpha_--1} }
            {\Gamma(\alpha_+)\,\Gamma(\alpha_-)\, (v-v_{_F})^{\alpha_+}\,(v+v_{_F})^{\alpha_-}}
            \\
     \times
     \Bigg\{
     H(\delta\om) \Int_{0}^{+\infty} w^{\alpha_+-1} (1+w)^{\alpha_--1} \dd w
     + H(-\delta\om) \Int_{0}^{+\infty}  w^{\alpha_--1} (1+w)^{\alpha_+-1} \dd w
     \Bigg\},
\end{multline}
where we have used $p'(\lambda)=2\pi\rho(\lambda)$. Note that we have already taken the $\delta\om\to 0$ limit inside of the remaining integrals, hence extending their integration domain up to $\infty$ (recall that  $\eps_{\delta\om}$ is such that $\eps_{\delta\om}/|\delta\om|\to +\infty$).
We stress that, since these limiting integrals are finite, such an operation does not affect the leading $\delta\om\to 0$ behavior of the overall expression.
These last two integrals correspond to representations of the beta function:
\begin{equation*}
   \Int_{0}^{+\infty}  w^{\alpha-1} (1+w)^{\beta-1} \dd w
   =\Int_0^1u^{\alpha-1}(1-u)^{-\alpha-\beta}\dd u
   =B(\alpha,1-\alpha-\beta)
   =\frac{\Gamma(\alpha)\,\Gamma(1-\alpha-\beta)}{\Gamma(1-\beta)}.
\end{equation*}
As a consequence, we obtain
\begin{multline}\label{TF-part4}
   \Int_{\Rset^2} \dd x\, \dd t \, e^{i(\om t-kx)}
     \moy{ \mathcal{O}^\dagger(x,t)\,\mathcal{O}(0,0) }
 = \frac{\Gamma(1-\alpha_+-\alpha_-)\,  \big|\mathcal{F}^{\mathcal{O}}_{{}_{1,-1}}\big|^2\,}
            {\pi (v-v_{_F})^{\alpha_+}\,(v+v_{_F})^{\alpha_-}}
            \\
     \times
     \Big\{
     H(\delta\om) \sin\pi\alpha_-
     + H(-\delta\om) \sin\pi\alpha_+
     \Big\}\, \left|\frac{\delta\om}{2\pi}\right|^{\alpha_++\alpha_--1}  +o(|\delta\om|^{\alpha_++\alpha_--1}).
\end{multline}
We recall that here $\alpha_+=(F_{1,-1}^+-1)^2$ and $\alpha_-=(F_{1,-1}^-)^2$.

\subsubsection{Behavior of the Fourier transform of the two-point function close to the hole excitation threshold}

We proceed similarly to obtain the behavior of the Fourier transform \eqref{TF-hole1} in the vicinity of the one-hole excitation threshold. We use again \eqref{result-int}  with $\alpha_+=(F_{-1,1}^+ +1)^2$, $\alpha_-=(F_{-1,1}^-)^2$, and $E=\delta\om+\nu\veps'(\la)$, $P=\nu p'(\la)$. Hence we have
\begin{equation}
  E \pm v_{_F}P=\delta\om+\nu\veps'(\la)\pm v_{_F} \nu p'(\la)
                           = \delta\om + \nu p'(\la) [ v\pm v_{_F} ].
\end{equation}
Note that, in the present case, $v<v_{_F}$,
and the two Heaviside functions restrict the domain of integration on $\nu$ such that
\begin{equation}\label{rest-nu-hole}
   -\frac{\delta\om}{p'(\la)(v_{_F}+v)}<\nu<\frac{\delta\om}{p'(\la)(v_{_F}-v)}.
\end{equation}
In particular, the domain of integration in not empty only if $\delta\om>0$.
Hence,
\begin{multline}\label{TF-hole2}
   \Int_{\Rset^2} \dd x\, \dd t \, e^{i(\om t-kx)}
     \moy{ \mathcal{O}^\dagger(x,t)\,\mathcal{O}(0,0) }_{k_h,\varepsilon_h+\delta\omega}
 = \frac{2\pi \rho(\la) \, \big|\mathcal{F}^{\mathcal{O}}_{{}_{-1,1}}\big|^2 }
            {[4\pi v_{_F}]^{\alpha_+ +\alpha_- -1}\,\Gamma(\alpha_+)\,\Gamma(\alpha_-)}
            \\
     \times
     H(\delta\om) \int_{-\frac{\delta\om}{p'(\la)[v_{_F}+v]}}^{\frac{\delta\om}{p'(\la)[v_{_F}-v]}}
        [ \delta\om + \nu p'(\la) ( v_{_F}+v )]^{\alpha_+ -1}\,
        [ \delta\om - \nu p'(\la) ( v_{_F}-v )]^{\alpha_- -1} \,\dd\nu.
\end{multline}
Changing  variables $w=(v_{_F}-v)[ \delta\om + \nu p'(\la) ( v_{_F}+v )]/(2v_{_F}\delta\om)$ in the integral, we reduce it to the integral representation of the beta function, and finally obtain
\begin{align}
   &\Int_{\Rset^2} \dd x\, \dd t \, e^{i(\om t-kx)}
     \moy{ \mathcal{O}^\dagger(x,t)\,\mathcal{O}(0,0) }_{k_h,\varepsilon_h+\delta\omega}
        \nonumber\\
 \qquad
 &\qquad= H(\delta\om)\,\frac{ \big|\mathcal{F}^{\mathcal{O}}_{{}_{-1,1}}\big|^2\,
 \left(\frac{\delta\om}{2\pi}\right)^{\alpha_++\alpha_--1} }
            {\Gamma(\alpha_+)\,\Gamma(\alpha_-)\, (v_{_F}-v)^{\alpha_+}\,(v_{_F}+v)^{\alpha_-}}
     \Int_{0}^{1} w^{\alpha_+-1} (1-w)^{\alpha_--1} \dd w
     \\
& \qquad
 = H(\delta\om)\, \frac{ \big|\mathcal{F}^{\mathcal{O}}_{{}_{-1,1}}\big|^2 }
            {\Gamma(\alpha_+ +\alpha_-)\, (v_{_F}-v)^{\alpha_+}\,(v_{_F}+v)^{\alpha_-}}     \;
      \left(\frac{\delta\om}{2\pi}\right)^{\alpha_++\alpha_--1}. \label{TF-hole4}
\end{align}
We recall that here $\alpha_+=(F_{-1,1}^+ +1)^2$ and $\alpha_-=(F_{-1,1}^-)^2$.

\subsection{Density structure factor}
\label{subsec-dens-strf}

We now turn to the computation of the density structure factor \eqref{DSF}.
As in Section~\ref{subsec-corr-dens}, it is convenient to introduce the generating function for the correlation function of densities (see formula \eqref{gen_f_deriv}):
\begin{align}
   S(k,\om)
    &= -\frac{1}{8\pi^2} \Int_{\Rset^2}\,\dd x\,\dd t\; e^{i(\om t- k x)}
            \frac{\partial^2}{\partial x^2}\frac{\partial^2}{\partial\alpha^2}
            \moy{ e^{it H}\, e^{2\pi i \alpha \mathcal{Q}(x)}\, e^{-it H_{\alpha}} } \Big|_{\alpha=0}
            \nonumber\\
    &= \frac{k^2}{8\pi^2} \Int_{\Rset^2}\,\dd x\,\dd t\; e^{i(\om t- k x)}
            \frac{\partial^2}{\partial\alpha^2}
            \moy{ e^{it H}\, e^{2\pi i \alpha \mathcal{Q}(x)}\, e^{-it H_{\alpha}} } \Big|_{\alpha=0}.
\end{align}
Summing over $\alpha$-twisted form factors of the type \eqref{Expan-GF}, we obtain
\begin{equation}
    S(k,\om)=  \frac{k^2\, \rho(\la)}{8\pi^2} \!
             \Int_{-\eps_{\delta\om}}^{\eps_{\delta\om}}\!\! \dd\nu   \Int_{\Rset^2}\! \dd x\, \dd t \,
             \frac{\partial^2}{\partial\alpha^2}
      \frac{|\mathcal{F}^\alpha_{\tau,-\tau}|^2\,
               e^{i x[2\alpha k_{_F}+\tau \nu p'(\la)]+it [\delta\om-\tau \nu \veps'(\la)] } }
           { [-2\pi i (x-v_{{}_F}t)] ^{\alpha_+^\tau(\alpha)}\,
              [ 2\pi i (x+v_{{}_F}t)]^{\alpha_-^\tau(\alpha)}   }  \Bigg|_{\alpha=0} \!  ,      \!
\end{equation}
where $\tau=+1$ for contributions around the particle threshold and $\tau=-1$ for contributions around the hole threshold.
The edge exponents  read $\alpha_\pm^\tau(\alpha)=[(\alpha-\tau)\mathcal{Z}+\tau\Phi_\pm]^2$, where $\mathcal{Z}$ is the value of the fractional charge on the Fermi boundary
and where  $\Phi_\pm$ are given by \eqref{Phi+}, \eqref{Phi-} upon the exchange $\la_0 \hookrightarrow  \la$.
We remind that  $|\mathcal{F}^\alpha_{\tau,-\tau}|^2$ is proportional to $\alpha^2$ (see Appendix~\ref{twist}). Hence, the second $\alpha$-derivative
has to act on this factor only, meaning that we can directly set $\alpha=0$ in the remaining part of the integrand.
Therefore, re-expressing $\partial_\alpha \big[ \mathcal{F}^\alpha_{\pm1,\mp1}  \big]_{\alpha=0}$ with the help of \eqref{Fa-densityFF} and applying the general scheme described above, we get that the leading power-law behavior of the density structure factor on the particle and hole thresholds is respectively given  by the following contributions:
\begin{multline}\label{str-f-p}
  S(k,\om)_{\text{part}}
 = \frac{\Gamma(1-\alpha_+-\alpha_-)\,  \big|\mathcal{F}^{j}_{{}_{1,-1}}\big|^2\,}
            {\pi (v-v_{_F})^{\alpha_+}\,(v+v_{_F})^{\alpha_-}}
            \\
     \times
     \Big\{
     H(\delta\om) \sin\pi\alpha_-
     + H(-\delta\om) \sin\pi\alpha_+
     \Big\}\, \left|\frac{\delta\om}{2\pi}\right|^{\alpha_++\alpha_--1},
\end{multline}
around the particle threshold, and
\begin{equation}\label{str-f-h}
  S(k,\om)_{\text{hole}}
 = H(\delta\om)\,\frac{  \big|\mathcal{F}^{j}_{{}_{-1,1}}\big|^2\,}
            {\Gamma(\alpha_+ +\alpha_-)\, (v-v_{_F})^{\alpha_+}\,(v+v_{_F})^{\alpha_-}} \,
     \left(\frac{\delta\om}{2\pi}\right)^{\alpha_++\alpha_--1},
\end{equation}
around the hole threshold.
In both expressions \eqref{str-f-p} and \eqref{str-f-h}, the edge exponents are equal to $\alpha_\pm=[\mathcal{Z}-\Phi_\pm]^2$.  They can be easily expressed in terms of the fractional charge and dressed phase,
\begin{align}
\alpha_+=&\left(\frac {\mathcal{Z}}2+ \frac 1{2\mathcal{Z}}+\phi(q,\la)\right)^2,\\
\alpha_-=&\left(\frac {\mathcal{Z}}2- \frac 1{2\mathcal{Z}}+\phi(-q,\la)\right)^2.
\end{align}
These results agree with the non-linear Luttinger liquid predictions \cite{ImaG09,ImaSG11}.

\subsection{Spectral function}
\label{subsec-spf}

Let us now consider the spectral function \eqref{spectr-fct}.
Using the spatial and temporal translation invariance of the model, we can recast the initial definition \eqref{spectr-fct} as:
\begin{equation}\label{spectr-bis}
  A(k,\om)=\frac{\text{sign}(\om)}{2\pi}\!  \Int_{\Rset^2}\! \dd x\, \dd t \Big\{ e^{i(\om t-k x)}
            \moy{\Psi(x,t)\,\Psi^\dagger(0,0)}
            + e^{-i(\om t-k x)} \moy{\Psi^\dagger(x,t)\,\Psi(0,0)} \Big\}.
\end{equation}
The limiting behavior of each of these two terms can easily be deduced from the general scheme described above.

The first term (which is non-zero only when $\om>0$) gives the leading power-law behavior of $A(k,\om)$ in the vicinity of $(k_p,\veps_p)$ (particle threshold) or of $(k_h,\veps_h)$ (hole threshold) as explained in Section~\ref{subsec-TF}.
We obtain the following result around the one-particle threshold:
\begin{multline}\label{A-part}
   A(k,\om)_{\text{part}} \; \simeq \; \frac{\Gamma(1-\alpha_+^{(p)}-\alpha_-^{(p)})\,  \big|\mathcal{F}^{\Psi^\dagger}_{{}_{1,-1}}\big|^2\,}
            {2\pi^2\, (v-v_{_F})^{\alpha_+^{(p)}}\,(v+v_{_F})^{\alpha_-^{(p)} }}
            \\
     \times
     \Big\{
     H(\delta\om) \sin\pi\alpha_-^{(p)}
     + H(-\delta\om) \sin\pi\alpha_+^{(p)}
     \Big\}\, \left|\frac{\delta\om}{2\pi}\right|^{\alpha_+^{(p)}+\alpha_-^{(p)}-1},
\end{multline}
with exponents
\begin{equation}
 \alpha_\pm^{(p)}=\left(\mathcal{Z}\mp\frac{1}{2\mathcal{Z}}-\Phi_\pm\right)^2=
\left(\frac {\mathcal{Z}}2+\phi(\pm q,\la)\right)^2.
\end{equation}
The result is slightly different around the one-hole threshold:
\begin{equation}\label{A-hole}
   A(k,\om)_{\text{hole}} \; \simeq \;
   \frac{H(\delta\om)\,\big|\mathcal{F}^{\Psi^\dagger}_{{}_{-1,1}}\big|^2 }
            {2\pi\Gamma(\alpha_+^{(h)} +\alpha_-^{(h)})\, (v_{_F}-v)^{\alpha_+^{(h)}}\,(v_{_F}+v)^{\alpha_-^{(h)}}}     \;
      \left(\frac{\delta\om}{2\pi}\right)^{\alpha_+^{(h)}+\alpha_-^{(h)}-1},
\end{equation}
with exponents
\begin{equation}
 \alpha_\pm^{(h)}=\left(\mathcal{Z}\pm\frac{1}{2\mathcal{Z}}-\Phi_\pm\right)^2=\left(\frac {\mathcal{Z}}2\pm\frac{1}{\mathcal{Z}}+\phi(\pm q,\la)\right)^2.
\end{equation}

The second term of \eqref{spectr-bis} (which is non-zero only for non-positive $\om$) corresponds to a Fourier transform of the type \eqref{loc-TF} evaluated at the point $(-k,-\om)$.
Hence, taking into account the symmetry $k\to -k$, the study of this term through the line of Section~\ref{subsec-TF} gives the leading power-law behavior of $A(k,\om)$ in the vicinity of $(k_p,-\veps_p)$ or of $(k_h,-\veps_h)$. We obtain respectively:
\begin{multline}\label{A-part-}
   A(k,\om)_{\overline{\text{part}}} \; \simeq \;  -\frac{\Gamma(1-\alpha_+^{(h)}-\alpha_-^{(h)})\,  \big|\mathcal{F}^{\Psi}_{{}_{1,-1}}\big|^2\,}
            { 2\pi^2\,(v-v_{_F})^{\alpha_+^{(h)}}\,(v+v_{_F})^{\alpha_-^{(h)} }}
            \\
     \times
     \Big\{
     H(-\om-\veps_p) \sin\pi\alpha_-^{(h)}
     + H(\om+\veps_p) \sin\pi\alpha_+^{(h)}
     \Big\}\, \left|\frac{\om+\veps_p}{2\pi}\right|^{\alpha_+^{(h)}+\alpha_-^{(h)}-1}
\end{multline}
for the leading power-law behavior around $(k_p,-\veps_p)$, and
\begin{equation}\label{A-hole-}
   A(k,\om)_{\overline{\text{hole}}}  \; \simeq \;
  -\frac{ H(-\om-\veps_h)\,  \big|\mathcal{F}^{\Psi}_{{}_{-1,1}}\big|^2 }
            {2\pi\Gamma(\alpha_+^{(p)} +\alpha_-^{(p)})\, (v_{_F}-v)^{\alpha_+^{(p)}} (v_{_F}+v)^{\alpha_-^{(p)}}}     \;
      \left|\frac{\om+\veps_h}{2\pi}\right|^{\alpha_+^{(p)}+\alpha_-^{(p)}-1}
\end{equation}
for the leading power-law behavior around $(k_h,-\veps_h)$.
Note that the edge exponent in \eqref{A-part-} (resp. in \eqref{A-hole-}) coincides with the edge exponent in \eqref{A-hole} (resp. in \eqref{A-part}).

\section{Conclusion}
\label{sec-concl}

To conclude this paper, it seems appropriate to outline once again the main features of our method before discussing possible further developments. In contrast to what happens in the case of massive models, one encounters serious difficulties when studying correlation functions from their form factor series in critical models.
As we have shown, a macroscopic description of the excited states is not enough for a successful summation of the form factor series, and
a formal replacement of the sum by an integral over the particle/hole rapidities leads to meaningless results.
This problem was already noted in the literature, see e.g.  \cite{LesSS96,LesS97,LesKS2003}.

In fact, the success of our method comes from the fact that it is  built on a microscopic description of the excited states. Such a microscopic description enables us to explicitly sum up the quasi-critical form factors belonging to a given class. This amounts to dress the original
bare form factor (taken for example as the simplest representative of the given class)  by the complete cloud of  excitations having zero momentum and energy in the thermodynamic limit. Then, the resulting dressed form factors admit a purely macroscopic description,
at least for  the examples that have been considered above. It shed some light on the relation between the particle description in the microscopic model and the effective one corresponding to its thermodynamic limit.

In the present paper, we have dealt with a situation with (contributing) excited states of particle/hole type only, and where the particles and holes separated from
the Fermi boundaries are localized in a small vicinity of a certain point. A possible direction for
further developments would be to apply this method to more general classes of form factors.
In particular, a very interesting example to consider is the XXZ Heisenberg chain, for which we also need to take into account the contribution of bound states (see e.g. \cite{Mah81,PerWA09}).

\section*{Acknowledgements}

K.K.K., J.M.M., N.A.S. and V.T. are supported by CNRS.
N.K, K.K.K., J.M.M. and V.T. are supported by ANR grant  ANR-10-BLAN-0120-04-DIADEMS. K. K. K. and N. K. are supported by the CNRS grant  PEPS-PTI - Asymptotique d'int\'egrales multiples.
N.K. is supported by the Burgundy region, FABER grant 2010-9201AAO047S00753.
We also acknowledge  the support from the GDRI-471 of CNRS `French-Russian network in Theoretical and Mathematical  Physics'.
N.A.S. is also supported by the Program of RAS Basic Problems of the Nonlinear Dynamics,
RFBR-11-01-00440, RFBR-11-01-12037-ofi-m, SS-4612.2012.1.
When this work was done, K.K.K. was supported by the EU Marie-Curie Excellence Grant MEXT-CT-2006-042695, DESY and IUPUI.
N.K., N.A.S. and K.K.K. would like to thank the Theoretical Physics group of the Laboratory of Physics at ENS Lyon for hospitality, which makes this collaboration possible.
N.K., J.M.M. and V.T. would  like to thank LPTHE (Paris VI University) for hospitality.

\appendix

\section{Thermodynamic limit of the model}
\label{app-therm}

In this appendix, we remind some definitions of thermodynamic quantities that are useful for our study.

We recall that, in the thermodynamic limit, the set of Bethe roots $\la_j$  \eqref{Bethe-gr} for the ground state densely fills
the Fermi zone $[-q,q]$ with a density  $\rho(\la)$. This function solves the following integral equation:
\begin{equation}\label{Lieb-eq}
   \rho(\la)-\frac{1}{2\pi}\Int_{-q}^q K(\la-\mu)\, \rho(\mu)\, \dd\mu = \frac{1}{2\pi},
   \qquad \text{with}\quad K(\la)=\theta'(\la)=\frac{2c}{\la^2+c^2}.
\end{equation}
The dressed momentum $p(\la)$ corresponds to the antiderivative of the density function that vanishes at the origin:
\begin{equation}\label{dmom}
  p(\la)=p_0(\la)+\Int_{-q}^q\theta(\la-\mu)\,\rho(\mu)\,\dd\mu=2\pi\Int_{0}^\la \rho(\mu)\,\dd\mu.
\end{equation}
The value of the Fermi boundary $q$ is determined by the dressed energy $\veps(\lambda)$. The latter
  is defined as the unique solution to the integral equation
\begin{equation}
   \veps(\la)-\frac{1}{2\pi}\Int_{-q}^q K(\la-\mu)\,\veps(\mu)\,\dd\mu=\veps_0(\la). \label{denergy}
\end{equation}
The parameter $q$ is then chosen in such a way that the unique solution to the above equation also verifies $\veps(\pm q)=0$.
The solvability of this non-linear problem has been established recently in \cite{Koz2012}. Note that the value of the
parameter $q$ then fixes throught $\int_{-q}^q\rho(\la)\dd\la=D$ the value $D=\lim (N/L)$ of the average density.

Other important thermodynamic quantities are also defined as solutions of linear integral equations. Among them, let us mention the dressed phase $\phi$ and dressed charge $Z$:
\begin{align}
   &\phi(\la,\mu)-\frac{1}{2\pi}\Int_{-q}^qK(\la-\omega)\, \phi(\omega,\mu)\, \dd\omega
      =\frac{\theta(\la-\mu)}{2\pi},\label{dphase}\\
   &Z(\la)-\frac{1}{2\pi}\Int_{-q}^qK(\la-\omega)\, Z(\omega)\,\dd\omega =1. \label{dcharge}
\end{align}
Evidently the dressed phase and the dressed charge are not independent functions. In particular, it
follows from \eqref{dphase}, \eqref{dcharge} that
\begin{equation}\label{Z-Phi1}
   Z(\lambda)=1+\phi(\lambda,-q)-\phi(\lambda,q).
 \end{equation}
There exists one more useful equation relating $\mathcal{Z}=Z(\pm q)$ and the dressed phase \cite{KorS98}:
\begin{equation}\label{Z-Phi2}
   \mathcal{Z}^{-1}=1-\phi(q,-q)-\phi(q,q)=1+\phi(-q,q)+\phi(-q,-q).
\end{equation}

We associate to each excited state  (defined through the Bethe equations \eqref{eq-Bethe})  a counting function $\xi(\omega)$. It realizes a one-to-one correspondence
between the Bethe roots $\widehat{\mu}_{\ell_j}$ and the integers $\ell_j$: $\xi(\widehat\mu_{\ell_j})=\ell_j/L$.
In finite volume, each counting function depends on the particular configuration of all integers $\ell_j$ labeling the corresponding state but, in the thermodynamic limit (up to
corrections in $1/L$), the counting functions of any particle/hole states (with $N'$ such that $N'-N$ remains finite) coincide. Their common limit is given by the following simple combination of the dressed
momentum $p$ and average density $D$:
\begin{equation}\label{xi}
   \underset{L\to\infty}{\lim}\xi(\om) = \frac{p(\om)}{2\pi} +\frac{D}{2}.
\end{equation}
For a given particle-hole excited state, the counting function allows one to define the particle and hole macroscopic rapidities $\widehat\mu_{p_a}$ and $\widehat\mu_{h_a}$ as the pre-image of the corresponding integers $p_a$ and $h_a$:
\begin{equation}\label{def-rap-ph}
  \widehat\mu_{p_a}=\xi^{-1}(p_a/L),\qquad \widehat\mu_{h_a}=\xi^{-1}(h_a/L),\qquad a=1,\ldots n.
\end{equation}

In general, the existence of particles and holes generates a global shift in the position of the Bethe roots of an excited state \eqref{eq-Bethe} with respect to those of the physical ground state \eqref{Bethe-gr}:
\begin{equation}\label{shift}
   \widehat{\mu}_a   -   \widehat{\la}_a
   = \frac{ F(\widehat\la_a)}
              {L\, \rho(\widehat\la_a)} + O(L^{-2}),
              \qquad
              a=1,\ldots,N, \quad a\neq h_1,\ldots, h_n.
\end{equation}
Here $F$ is the shift function, which depends in fact on the macroscopic rapidities $\mu_{p_a}$ and $\mu_{h_a}$
associated to the positions of the particles and holes:
\begin{equation}
   F(\la)\equiv F\bigg(\la \bigg| \begin{matrix} \{\mu_{p_j}\} \\ \{\mu_{h_j}\} \end{matrix} \bigg).
\end{equation}
It is defined as the solution of the linear integral equation
\begin{equation}
  F(\la)  -\Int_{-q}^q\! K(\la-\mu)\,   F(\mu)  \, \frac{\dd\mu}{2\pi}
  = -\frac{\Delta N}{2} \bigg[1+\frac{\theta(\la-q)}{\pi}\bigg]
  -\sum_{k=1}^n\frac{\theta(\la-\mu_{p_k})-\theta(\la-\mu_{h_k})}{2\pi} \; .
  \label{eq-shift}
\end{equation}
It can be expressed, in terms of the dressed phase \eqref{dphase} and dressed charge \eqref{dcharge}, as
\begin{equation}
   F(\la)= -\Delta N \bigg[\frac{Z(\la)}{2}+\phi(\la,q)\bigg]-\sum_{a=1}^n \big[ \phi(\la,\mu_{p_a})-\phi(\la,\mu_{h_a})\big], \label{shift-ph}
\end{equation}
with $\Delta N= N'-N$ being finite in the thermodynamic limit.

\section{Form factor expansion for the correlation function of densities \label{twist}}

It was shown in  \cite{KitKMST07} that
\begin{equation}\label{Exp-val-Q-c}
 \frac{\bra{\psi'_\alpha(\{\widehat\mu\}) }\, e^{-2\pi i\alpha\mathcal{Q}(x)} \ket{\psi_g } }
         {{\|\psi'_\alpha(\{\widehat\mu\})\| \, \|\psi_g\|}}
 =
 e^{-ix\mathcal{P}_{\mathrm{ex}}}\, \mathcal{F}^\alpha(\{\widehat\mu\}),
\end{equation}
for an arbitrary state  $\ket{ \psi'_\alpha(\{\widehat\mu\}) }$ depending on parameters $\{\widehat\mu\}$ satisfying the $\alpha$-twisted Bethe equations \eqref{eq-Bethe-alpha}.
Here $\mathcal{F}^\alpha(\{\widehat\mu\})$ is given by \eqref{def-F}, and $\mathcal{P}_{\mathrm{ex}}=\sum_{j=1}^N(\widehat\mu_{\ell_j}-\widehat\lambda_j)$.
If $\alpha$ is a real parameter, then all $\widehat\mu_{\ell_k}$ are real, and taking the hermitian conjugation of \eqref{Exp-val-Q1} we obtain
 \begin{equation}\label{Exp-val-Q1}
    \frac{\bra{\psi_g} \, e^{2\pi i\alpha\mathcal{Q}(x)}  \ket{\psi'_\alpha(\{\widehat\mu\})} }
           {{\|\psi'_\alpha(\{\widehat\mu\})\| \, \|\psi_g\|}}
    =
    e^{ix\mathcal{P}_{\mathrm{ex}}} \, \mathcal{F}^\alpha(\{\widehat\mu\}).
\end{equation}

Let us act with the operator $\partial_x\partial_\alpha$ at $\alpha=0$ on the l.h.s. of \eqref{Exp-val-Q1}.   Using $\partial_x\mathcal{Q}(x)=j(x,0)$ we have
\begin{equation}\label{DlExp-val-Q}
   \partial_x\partial_\alpha \! \left.
   \frac{\bra{\psi_g}\, e^{2\pi i\alpha\mathcal{Q}(x)}  \ket{\psi'_\alpha(\{\widehat\mu\}) } }
           {{\|\psi'_\alpha(\{\widehat\mu\})\| \, \|\psi_g\|}}
   \right|_{\alpha=0}
   =2\pi i\frac{\bra{\psi_g}\, j(x,0) \, \ket{\psi'(\{\widehat\mu\})} }
                     {{\|\psi'(\{\widehat\mu\})\| \,  \|\psi_g\|}},
 \end{equation}
where $\ket{\psi'(\{\widehat\mu\})}=\left. \ket{\psi'_\alpha(\{\widehat\mu\})}\right|_{\alpha=0}$.
On the other hand, applying the same differential operator to the r.h.s. of \eqref{Exp-val-Q1} we obtain
\begin{equation}\label{DrExp-val-Q}
   \partial_x\partial_\alpha\! \left.
   \frac{\bra{\psi_g}\, e^{2\pi i\alpha\mathcal{Q}(x)}  \ket{\psi'_\alpha(\{\widehat\mu\}) } }
           {{\|\psi'_\alpha(\{\widehat\mu\})\| \, \|\psi_g\|}}
   \right|_{\alpha=0}
   =2\pi i\delta_{\psi',\psi_g} D
 +i(1-\delta_{\psi',\psi_g})\mathcal{P}_{\mathrm{ex}}\, e^{ix\mathcal{P}_{\mathrm{ex}}} \,
   \partial_\alpha \mathcal{F}^\alpha(\{\widehat\mu\})\big|_{\alpha=0}.
\end{equation}
Here $\delta_{\psi',\psi_g}=1$ if $\{\widehat\mu\}=\{\widehat\lambda\}$ at $\alpha=0$ and $\delta_{\psi',\psi_g}=0$ otherwise.
We also have used that $\mathcal{P}_{\mathrm{ex}}=2\pi\alpha D$ for the  state with $\ell_j=j$ (see \eqref{eq-Bethe-alpha}).
Comparing \eqref{DlExp-val-Q} and \eqref{DrExp-val-Q} we find that
\begin{equation}\label{FF-j-1}
 \frac{\bra{\psi_g} \, j(x,0) \, \ket{\psi'(\{\widehat\mu\})}}{{\|\psi'(\{\widehat\mu\})\| \, \|\psi_g\|}}
 =\delta_{\psi',\psi_g} D
 +\frac{(1-\delta_{\psi',\psi_g})}{2\pi} \, \mathcal{P}_{\mathrm{ex}} \, e^{ix\mathcal{P}_{\mathrm{ex}}} \,
   \partial_\alpha \mathcal{F}^\alpha(\{\widehat\mu\})\big|_{\alpha=0}.
 \end{equation}
Similarly the equation \eqref{Exp-val-Q-c} leads to the following representation:
\begin{equation}\label{FF-j-2}
 \frac{\bra{\psi'(\{\widehat\mu\})} \, j(x,0)\, \ket{\psi_g}}{{\|\psi'(\{\widehat\mu\})\| \, \|\psi_g\|}}
 =\delta_{\psi',\psi_g} D
 +\frac{(1-\delta_{\psi',\psi_g})}{2\pi} \, \mathcal{P}_{\mathrm{ex}} \, e^{-ix\mathcal{P}_{\mathrm{ex}}} \,
   \partial_\alpha \mathcal{F}^\alpha(\{\widehat\mu\})\big|_{\alpha=0}.
\end{equation}

Using these formulas we can prove  \eqref{gen_f_deriv}.
On the one hand, we have
\begin{equation}\label{Expan-1}
  \moy{ j(x,t) \, j(0,0)}
 =\sum_{|\psi'(\{\widehat\mu\})\rangle}
   e^{-it\mathcal{E}_{\mathrm{ex}}}
   \frac{\bra{\psi_g}\, j(x,0) \, \ket{\psi'(\{\widehat\mu\}) }
            \bra{\psi'(\{\widehat\mu\})} \, j(0,0) \, \ket{\psi_g} }{\|\psi'(\{\widehat\mu\}) \|^2 \, \|\psi_g\|^2} ,
\end{equation}
where the sum is taken over the eigenstates of the Hamiltonian with periodic boundary conditions, which means that the parameters $\widehat\mu_{\ell_k}$  satisfy the equations \eqref{eq-Bethe-alpha} with $\alpha=0$.
Substituting \eqref{FF-j-2} and \eqref{FF-j-1} into \eqref{Expan-1}, we obtain
\begin{equation}\label{Expan-2}
   \moy{ j(x,t) \, j(0,0) } = D^2
 +\frac1{4\pi^2}\sum_{|\psi'(\{\widehat\mu\})\rangle\ne|\psi_g\rangle}
   \hspace{-2mm} {\mathcal{P}}_{\mathrm{ex}}^2 \,
   e^{-i t{\mathcal{E}}_{\mathrm{ex}} +ix {\mathcal{P}}_{\mathrm{ex}}  } \,
   \bigl( \partial_\alpha \mathcal{F}^\alpha(\{\widehat\mu\})\big|_{\alpha=0}\bigr)^2.
\end{equation}
On the other hand, one can expand the generating function
$\moy{ e^{itH}e^{2\pi i\alpha{\cal Q}(x)}  e^{-itH_\alpha} }$ as
\begin{equation}\label{Expan-Q}
   \moy{ e^{itH}e^{2\pi i\alpha{\cal Q}(x)}  e^{-itH_\alpha} }
   =
   \!\sum_{|\psi'_\alpha(\{\widehat\mu\})\rangle} \hspace{-2mm}
   \frac{\bra{\psi_g} \, e^{itH}e^{2\pi i\alpha{\cal Q}(x)}  e^{-itH_\alpha} \ket{\psi'_\alpha(\{\widehat\mu\}) }
            \moy{\psi'_\alpha(\{\widehat\mu\}) \, |\, \psi_g } }{\|{\psi'_\alpha(\{\widehat\mu\}) }\|^2  \, \|\psi_g\|^2},
\end{equation}
where the sum runs over the complete set of twisted eigenstates $\ket{\psi'_\alpha(\{\widehat\mu\}) }$.
Using \eqref{Exp-val-Q1}, we obtain
\begin{equation}\label{Expan-Q1}
 \moy{ e^{itH}e^{2\pi i\alpha{\cal Q}(x)}  e^{-itH_\alpha} }
 =
 \!\sum_{|\psi'_\alpha(\{\widehat\mu\})\rangle} \hspace{-1mm}
 e^{-i t{\mathcal{E}}_{\mathrm{ex}} +ix {\mathcal{P}}_{\mathrm{ex}}  } \,
 \bigl( \mathcal{F}^\alpha(\{\widehat\mu\})\bigr)^2.
\end{equation}
Note that, in \eqref{Expan-Q1}, ${\mathcal{E}}_{\mathrm{ex}} $ and ${\mathcal{P}}_{\mathrm{ex}} $ are both given in terms of the roots $\widehat\mu_{\ell_k}$ of the twisted Bethe equations  \eqref{eq-Bethe-alpha}.
It is easy to see that, acting with the operator $-\partial^2_x\partial^2_\alpha/8\pi^2$ at $\alpha=0$ on the equation \eqref{Expan-Q1}, we obtain \eqref{Expan-2}.
Indeed,
\begin{equation}\label{Ortho}
   \mathcal{F}^\alpha(\{\widehat\mu\})\big|_{\alpha=0}=0,\quad\mbox{if}\quad
   \ket{\psi'_\alpha(\{\widehat\mu\})} \big|_{\alpha=0} \ne \ket{\psi_g},
\end{equation}
since the scalar product of two different eigenstates vanishes.
Hence, for the corresponding terms in the sum \eqref{Expan-Q1}, the second $\alpha$-derivative acts only on $\bigl(\mathcal{F}^\alpha(\{\widehat\mu\})\bigr)^2$, while in the rest of the formula one can set $\alpha=0$.
As for the action of the differential operator on the  term with $\ket{\psi'_\alpha(\{\widehat\mu\})}\big|_{\alpha=0}=\ket{\psi_g}$,
it gives $D^2$, since in this case $\mathcal{P}_{\mathrm{ex}}=2\pi\alpha D$.

Let us now express, for $\ket{\psi'_\alpha(\{\widehat\mu\})} \big|_{\alpha=0}\ne\ket{\psi_g}$, the derivative $\partial_\alpha \mathcal{F}^\alpha(\{\widehat\mu\})\big|_{\alpha=0}$ in terms of the form factor of the density  operator.
Differentiating \eqref{Exp-val-Q1} with respect to $\alpha$ at $\alpha=0$ and using \eqref{Ortho} we obtain
\begin{equation}\label{Alpha-der1}
 \left( e^{ix {\mathcal{P}}_{\mathrm{ex}}}-1\right)\,
 \partial_\alpha\mathcal{F}^\alpha(\{\widehat\mu\})\big|_{\alpha=0}
 =
 2\pi i\frac{\bra{\psi_g}\, \mathcal{Q}(x) \, \ket{\psi'(\{\widehat\mu\})} }
                 {{\|\psi'(\{\widehat\mu\})\| \, \|\psi_g\|}}.
\end{equation}
In particular, if $\ket{\psi'(\{\widehat\mu\}) }=\ket{\psi'_{\tau n_0,\ell} }$, then
\begin{equation}\label{Alpha-der2}
\left( e^{ix {\mathcal{P}}_{\mathrm{ex}}}-1\right)\,
 \partial_\alpha\mathcal{F}^{\alpha}_{{\tau n_0,\ell}}\big|_{\alpha=0}
 =
 2\pi i \; \mathcal{F}^{\mathcal{Q}(x)}_{{\tau n_0,\ell}} \, .
\end{equation}
Taking the derivative over $x$ at $x=0$, we finally get,
\begin{equation}\label{Fa-densityFF}
   \partial_\alpha\mathcal{F}^\alpha_{\tau n_0,\ell}\big|_{\alpha=0}
   =
 \frac{2\pi}{\mathcal{P}_{\mathrm{ex}}} \; \mathcal{F}^{j}_{\tau n_0,\ell} \, .
 \end{equation}
%


%

\end{document}